\documentclass[aps,pra,10pt,twocolumn,notitlepage,floatfix,superscriptaddress]{revtex4-2}
\usepackage[T1]{fontenc}
\usepackage[colorlinks=true,urlcolor=blue,linkcolor=blue,citecolor=blue]{hyperref}
\usepackage{graphicx,url,color,bbm,soul,mathtools,amssymb,amsmath,dsfont,lmodern,booktabs,mathrsfs,tikz,quantikz}

\newcommand{\QFT}{\mathcal{F}}
\newcommand{\im}{\text{i}}
\newcommand{\e}{\text{e}}
\newcommand{\diff}{\text{d}}

\begin{document}

\title{Quantum circuits for partial differential equations in Fourier space}

\author{Michael Lubasch}
\email{michael.lubasch@quantinuum.com}
\affiliation{Quantinuum, Partnership House, Carlisle Place, London SW1P 1BX, United Kingdom}

\author{Yuta Kikuchi}
\affiliation{Quantinuum K.K., Otemachi Financial City Grand Cube 3F, 1-9-2 Otemachi, Chiyoda-ku, Tokyo, Japan}
\affiliation{RIKEN Center for Interdisciplinary Theoretical and Mathematical Sciences (iTHEMS), RIKEN, Wako, Saitama 351-0198, Japan}

\author{Lewis Wright}
\affiliation{Quantinuum, Partnership House, Carlisle Place, London SW1P 1BX, United Kingdom}

\author{Conor Mc Keever}
\affiliation{Quantinuum, Partnership House, Carlisle Place, London SW1P 1BX, United Kingdom}

\date{December 22, 2025}

\begin{abstract}
For the solution of partial differential equations (PDEs), we show that the quantum Fourier transform (QFT) can enable the design of quantum circuits that are particularly simple, both conceptually and with regard to hardware requirements.
This is shown by explicit circuit constructions for the incompressible advection, heat, isotropic acoustic wave, and Poisson's equations as canonical examples.
We utilize quantum singular value transformation to develop circuits that are expected to be of optimal computational complexity.
Additionally, we consider approximations suited for smooth initial conditions and describe circuits that make lower demands on hardware.
The simple QFT-based circuits are efficient with respect to dimensionality and pave the way for current quantum computers to solve high-dimensional PDEs.
\end{abstract}

\maketitle

\section{Introduction}

Today's quantum hardware offers the exciting possibility to experimentally explore tasks for which quantum computers can be exponentially more efficient than classical computers.
One such task is the solution of high-dimensional partial differential equations (PDEs)~\cite{Be14, BeEtAl17, LlEtAl20, LiEtAl21, ChLiOs21}, of relevance to various scientific disciplines including computational fluid dynamics~\cite{We09}, structural mechanics~\cite{ZiTaFo14}, quantitative finance~\cite{Du06}, and quantum chemistry~\cite{Ja23}.
Although numerous quantum algorithms for solving PDEs exist~\cite{CaEtAl13, Be14, MoPa16, BeEtAl17, XuEtAl18, CoJoOs19, XuEtAl19, ArEtAl19, ChLi20, GiEtAl20, Ga20, LlEtAl20, Bu21, LiEtAl21, ChLiOs21, OzEtAl22, LiMoSh22, JiLiYu22, Kr23, JiLiYu23a, JiLiYu23b, AnLiLi23, SuEtAl23, BhSr23, ItSrSu24, TeMa24, JiLiYu24a, SaEtAl24b, BrLa24, InEtAl24, JiLi24, GoEtAl24, BhSr25, HsEtAl25, DeDa25}, their realization is challenging for the present generation of quantum processors that provide restricted qubit counts and lack quantum error correction.
In pursuit of more feasible alternatives for current quantum computers, several variational quantum algorithms for PDEs have been proposed~\cite{LuEtAl20, KyPaEl21, LiuEtAl21, SaEtAl21, AlEtAl22, PfHeSc22, DeEtAl22, LiEtAl22, PoEtAl22, JaEtAl23, GuEtAl23, AlKa23, LiWaFe23, SuEtAl23, PfHeSc23, LiEtAl24, PoEtAl24, SaEtAl24, InEtAl24, BhSr25, SaEtAl25, KoEtAl25}.
These variational quantum algorithmic approaches, however, typically have no theoretical convergence guarantees and they can suffer from large sampling costs when so-called barren plateaus~\cite{McEtAl18, CeEtAl21, UvBi21, ZhGa21, PaEtAl21, OrKiWi21, WaEtAl21, HoEtAl22, LaEtAl22, CePlLu23} are encountered during the variational optimization.
Therefore, in the context of PDEs, it is still an open question whether current quantum hardware can demonstrate an exponential quantum advantage.
This motivates renewed focus on developing quantum circuits that retain clear performance guarantees.

\begin{table}
\centering
\begin{ruledtabular}
\begin{tabular}{lll}
 &PDE & circuit depth \\
 \midrule
 (a)&incompressible advection & $O\left(n \min[N, t N + \log(1/\epsilon)]\right)$ \\
 &heat & $O\left(n N \min\left[1, \sqrt{t} \log(1/\epsilon)\right]\right)$ \\
 &isotropic acoustic wave & $\tilde{O}\left(d n \left[t N + \log(1/\epsilon)\right] \right)$ \\
 &Poisson's & $\tilde{O}\left(d^{2} N^{2} / \epsilon\right)$ \\
 \midrule
 (b)&incompressible advection & $O(1)$ \\
 &heat & $O\left(n \min\left[1, \sqrt{t} \log(1/\epsilon)\right]\right)$ \\
 &isotropic acoustic wave & $\tilde{O}\left(d^{3} k_{\text{max}}^{2} n t^{2} / \epsilon\right)$ \\
 &Poisson's & $\tilde{O}\left(d^{2} k_{\text{max}}^{2} n^{2} / \epsilon\right)$ \\
\end{tabular}
\end{ruledtabular}
\caption{
\label{tab:SummaryCircuitDepths}
Summary of the shallowest circuits developed in this paper that prepare an amplitude-encoded solution to the listed PDE.
Circuit depths listed in part (a) make no assumptions about the functions describing the initial conditions of the time-dependent PDEs or the source term of Poisson's equation, whereas those listed in (b) assume that these functions are smooth.
We consider the solution on a grid of $N = 2^{n}$ grid points per dimension and encoded in the amplitudes of a quantum circuit with $n$ qubits for each dimension.
We denote the total integration time by $t$, the desired error by $\epsilon$, the dimensionality of the studied equation by $d$, the maximum wavenumber required to represent the smooth function by $k_{\text{max}}$, and $\tilde{O}(\cdot)$ is $O(\cdot)$ without polylogarithmic factors.
All reported circuit depths exclude both the known depth of the QFT and the problem-specific depths of the state preparation circuits for the initial conditions or the source term.
Details on the associated ancilla qubit counts, postselection probabilities, and additional circuit constructions are provided in the respective sections.
}
\end{table}

In this paper, we contribute to the development of non-variational quantum algorithms for PDEs and present explicit circuits of simple structure with well-defined computational complexity.
Notably, most of our circuits are composed of one- and two-qubit gates only.
A key ingredient of our approach is the quantum Fourier transform (QFT)~\cite{Co02, NiCh10}, which we utilize to diagonalize the differential operators appearing in PDEs.
We describe two types of quantum circuit constructions for elementary variants of the advection, heat, wave, and Poisson's equations.
Firstly, using quantum singular value transformation (QSVT)~\cite{GiEtAl19, MaEtAl21} we derive quantum circuits that are anticipated to be of optimal asymptotic gate complexity.
Table~\ref{tab:SummaryCircuitDepths} (a) summarizes the corresponding results.
The QSVT circuit depths grow at least linearly with the number of grid points $N = 2^{n}$ per dimension where $n$ is the qubit count per dimension, which can be challenging for current quantum computers when $n$ is large.
Secondly, and for the reason mentioned above, we also analyze the PDEs on the assumption of smooth initial conditions, for which we formulate circuits that do not have this potentially large factor $N$ in their circuit depth scalings, as can be seen in Table~\ref{tab:SummaryCircuitDepths} (b).
All proposed circuits have depths that increase at most polynomially with the PDE dimensionality $d$.

Therefore, provided that the initial states of the time-dependent PDEs studied in this paper and the source term of Poisson's equation have efficient quantum circuit encodings, on a quantum computer we obtain efficient encodings for the the entire solutions of the considered PDEs even in high dimensions, whereas on a classical computer the same encodings would generally be inefficient and characterized by an exponential cost with respect to dimensionality.
In classical computing, the construction of the initial state or source term typically entails calculating and storing $N^{d}$ complex numbers, which has a classical computational complexity of $\Omega\left(N^{d}\right)$.
Then the application of the classical Fourier transform to the $N^{d}$ data points requires $O\left[d N^{d} \log(N)\right]$ operations in general~\cite{CoTu65, FrJo05}.
In quantum computing, the corresponding Fourier transform is realized by a product of $d$ independent QFTs parallely run in a total time scaling as $O\left[\log^{2}(N)\right]$.
Next, the quantum computations continue by applying our efficient circuits of depths summarized in Tab.~\ref{tab:SummaryCircuitDepths}, which are then followed by a product of $d$ independent inverse QFTs that has time complexity $O\left[\log^{2}(N)\right]$.
We note that, while the quantum circuit encoding of the initial state or source term is not efficient for generic functions~\cite{PlBr11, ZhLiYu22, SunEtAl23}, efficient quantum circuit representations have been found for many practically relevant functions~\cite{LuEtAl20, PlEtAl22, Ra22, AkEtAl23, MaGoSa23, MoEtAl24, RoEtAl25}.

Our paper is inspired by important recent discoveries relating to explicit quantum circuit solutions for PDEs~\cite{SaEtAl24a, WrEtAl24, HuEtAl24, BhSr24, GuHuLi24, JiLiYu24b, KhEtAl24, LiPaCa25, PfEtAl25}.
Compared with previous works, the main novelties of our article are that we exploit the QFT together with QSVT in all of our circuit designs and investigate the situation of smooth initial conditions.
However, our approach is restricted to simple PDE instances for which the QFT provides a diagonalizing basis, such as the specific linear PDEs considered here, whereas some of the other strategies can be applied to more general PDEs.
Closely related to our paper are also the seminal articles~\cite{CoJoOs19, ChLiOs21}.
Reference~\cite{CoJoOs19} contains quantum algorithms simulating the wave equation for more general boundary conditions than are possible using our approach; see~\cite{SuStCa21} for corresponding circuit constructions.
In~\cite{ChLiOs21} the authors present quantum algorithms to solve certain PDEs, including Poisson's equation, with quantum linear systems algorithms~\cite{ChKoSo17} that are asymptotically more efficient in achieving high precision than our methods.

The paper is organized as follows.
Section~\ref{sec:Background} contains descriptions of the conventions and central tools that are being used in this work.
We derive the quantum circuits to solve the incompressible advection, heat, isotropic acoustic wave, and Poisson's equations in Sec.~\ref{sec:AdvectionEquation}, Sec.~\ref{sec:HeatEquation}, Sec.~\ref{sec:WaveEquation}, and Sec.~\ref{sec:PoissonEquation}, respectively.
In Sec.~\ref{sec:Discussion} we present a concluding discussion and topics for future research.

\section{Background}
\label{sec:Background}

In the following, we explain the notation and main building blocks of the paper, such as the shifted QFT, finite-difference representations of derivatives, ways to implement various boundary conditions, and our QSVT conventions.

\begin{figure*}
\centering
\includegraphics{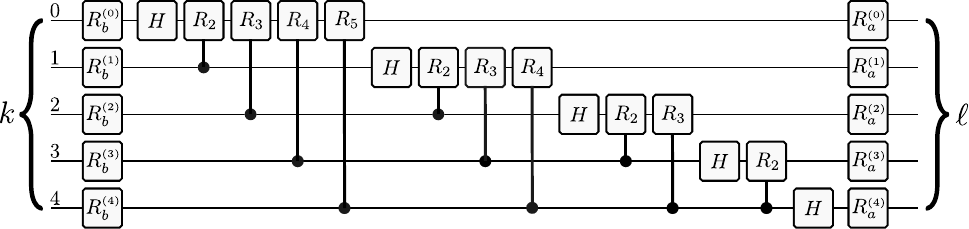}
\caption{\label{fig:ShiftedQFT}
Five-qubit circuit representation of the operator $\frac{1}{\sqrt{N}} \sum_{\ell} \e^{\im 2 \pi (k+a) (\ell+b) / N} \ket{\ell}\bra{k}$ realizing the QFT~\cite{Co02, NiCh10} with wavenumbers $k$ shifted by a constant $a$ and positions $\ell$ shifted by a constant $b$.
More specifically, the circuit implements the shifted QFT without the global phase $\e^{\im 2 \pi a b / N}$.
Here $n = 5$, $N = 2^{n}$, $R_{b}^{(\zeta)} = \text{diag}[1, \exp(\im \pi 2^{-\zeta} b)]$, $H$ is the Hadamard gate, $R_{\zeta} = \text{diag}[1, \exp(\im \pi 2^{-\zeta+1})]$ and $R_{a}^{(\zeta)} = \text{diag}[1, \exp(\im \pi 2^{\zeta-n+1} a)]$, where $\text{diag}[\cdot]$ denotes the diagonal single-qubit operator with $[\cdot]$ on the diagonal.
}
\end{figure*}

We assume that any $d$-dimensional, continuous function $f(x_{1}, x_{2}, \ldots, x_{d})$ is defined on the interval $(-1/2, 1/2)^{d}$ and satisfies periodic boundary conditions.
The function is discretized into a vector $|f\rangle$ of function values on the grid of $N^{d} = 2^{nd}$ grid points which in each dimension are chosen to be symmetric around $0$, i.e.\ $x_{\alpha} \in \{-1/2+1/(2N), -1/2+3/(2N), \ldots, 1/2-1/(2N)\}$ for all $\alpha \in \{1, 2, \ldots, d\}$.
We store $|f\rangle$ in the wave function coefficients corresponding to a qubit register of $nd$ qubits such that every dimension is represented by $n$ qubits and discretized using $N = 2^{n}$ grid points.
In each dimension $\alpha$, the grid point $x_{\alpha}^{(\ell)} = -1/2 + 1/(2N) + \ell/N$ for $\ell \in \{0, 1, \ldots, N-1\}$ is identified with the $n$-qubit computational basis state $\ket{\ell} = \ket{\ell_{0}, \ell_{1}, \ldots, \ell_{n-1}} = \ket{\ell_{0}} \otimes \ket{\ell_{1}} \otimes \ldots \otimes \ket{\ell_{n-1}}$ where $\ell_{\beta} \in \{0, 1\}$ represents the computational basis state of qubit $\beta$ and $\ell = \sum_{\beta = 0}^{n-1} 2^{\beta} \ell_{\beta}$.
Note that, in our binary encoding of $\ell$, qubit index $n-1$ corresponds to the most significant qubit.

Figure~\ref{fig:ShiftedQFT} shows a circuit corresponding to the QFT~\cite{Co02, NiCh10} with shifts for one dimension which we denote by $\mathcal{F}$ throughout the paper.
The exact QFT, which is illustrated in Fig.~\ref{fig:ShiftedQFT}, has a circuit depth scaling as $O\left(n^{2}\right)$.
By removing all rotation gates with angles smaller than a certain value from the exact QFT, one obtains the so-called approximate QFT~\cite{Co02, BaEtAl96} that has a circuit depth scaling as $O\left[n \log(n / \epsilon)\right]$ for error $\epsilon$ with respect to the exact QFT.
In our definition of the QFT we use input basis states $\ket{k} = \ket{k_{0}, k_{1}, \ldots, k_{n-1}}$ where $k_{\beta} \in \{0, 1\}$ and $k = \sum_{\beta = 0}^{n-1} 2^{n-1-\beta} k_{\beta}$, i.e.\ qubit $0$ is the most significant qubit.
The output basis states $|\ell\rangle$ are defined in the preceding paragraph.
We choose the shifts $a = -N/2$ and $b = -(N-1)/2$.
Then the shifted QFT has the entries
\begin{eqnarray}
 \bra{\ell} \QFT \ket{k} 
 = \frac{1}{\sqrt{N}} \e^{\im 2 \pi \left(k - \frac{N}{2}\right) \left(\frac{\ell}{N} - \frac{N-1}{2N}\right)}
 = \frac{1}{\sqrt{N}} \e^{\im 2 \pi \tilde{k} \tilde{\ell}}
\end{eqnarray}
where $\tilde{k} = k-N/2  \in \{-N/2, -N/2+1, \ldots, N/2-1\}$ and $\tilde{\ell} = -1/2+1/(2N)+\ell/N  \in \{-1/2+1/(2N), -1/2+3/(2N), \ldots, 1/2-1/(2N)\}$.
Therefore $\QFT$ contains in column $k$ the plane wave $w_{\tilde{k}}(x) = \e^{\im2\pi\tilde{k}x}/\sqrt{N}$ of wavenumber $\tilde{k}$ discretized via a grid of $N$ equidistant grid points symmetric around $0$.
These plane waves are a suitable basis for periodic boundary conditions, and the grid choice is consistent with our definition of discretized functions.
The generalization from one to $d$ dimensions is straightforward:
In $d$ dimensions we use $\QFT^{\otimes d} = \QFT_{1} \otimes \QFT_{2} \otimes \ldots \otimes \QFT_{d}$ where $\QFT_{\alpha} = \QFT$ for all $\alpha$ and we denote by $\tilde{k}_{\alpha}$ a wavenumber in dimension $\alpha$.

The operator representation $\hat{k}$ for wavenumber $\tilde{k}$ is
\begin{align}\label{eq:HatK}
\begin{split}
 \hat{k} 
 &= \sum_{\beta = 0}^{n-1} 2^{n-1-\beta} \left( \frac{\mathds{1}_{\beta}-Z_{\beta}}{2} \right) - \frac{N}{2}\mathds{1}
 \\
 &= -\frac{N}{4}\sum_{\beta = 0}^{n-1} 2^{-\beta} Z_{\beta} - \frac{1}{2}\mathds{1},
\end{split}
\end{align}
where $\mathds{1}$ is the global identity operator, $\mathds{1}_{\beta}$ the identity operator for qubit $\beta$ and $Z_{\beta}$ the Pauli $Z$ matrix acting on qubit $\beta$.

We realize the first derivative by means of the central difference approximation,
\begin{align}\label{eq:CentralDifference}
 \frac{\partial f}{\partial x_{\alpha}} \bigg|_{x_{\alpha}^{(\ell)}}
 \approx \frac{f\left(x_{\alpha}^{(\ell)}+s\right) - f\left(x_{\alpha}^{(\ell)}-s\right)}{2s},
\end{align}
where in our notation for $f$ we omit fixed variables and $s = 1/N$ is the grid spacing.
Note that this approximation of the first derivative for sufficiently small $s$ has the leading error of $O\left(s^{2}\right) = O\left(2^{-2n}\right)$ which can be systematically reduced by increasing $n$.
The plane wave $w_{\tilde{k}}(x)$ is an eigenfunction of the central difference operator in Eq.~\eqref{eq:CentralDifference} with eigenvalue $\im N \sin\left(2 \pi \tilde{k} / N\right)$:
\begin{align}\label{eq:CentralDifferencePlaneWaves}
\begin{split}
 \frac{\partial w_{\tilde{k}}}{\partial x} \bigg|_{\tilde{\ell}} 
 &\approx N \frac{\e^{\im 2 \pi \tilde{k} (\tilde{\ell}+1/N)} - \e^{\im 2 \pi \tilde{k} (\tilde{\ell}-1/N)}}{2 \sqrt{N}}
 \\
 &= \im N \sin\left(\frac{2 \pi \tilde{k}}{N}\right) w_{\tilde{k}}(\tilde{\ell}).
\end{split}
\end{align}
Therefore, we know the spectral decomposition of the central difference operator:
\begin{align}\label{eq:FirstDerivative}
 \frac{\partial}{\partial x}
 \ \widehat{=}\ 
 \im N \QFT \sin\left(\frac{2 \pi \hat{k}}{N}\right) \QFT^\dag.
\end{align}

In a similar way, we represent the second derivative by its finite difference approximation,
\begin{align}\label{eq:FiniteDifferenceLaplace}
 \frac{\partial^{2} f}{\partial x_{\alpha}^{2}} \bigg|_{x_{\alpha}^{(\ell)}}
 \approx 
 \frac{f\left(x_{\alpha}^{(\ell)}+s\right) - 2 f\left(x_{\alpha}^{(\ell)}\right) + f\left(x_{\alpha}^{(\ell)}-s\right)}{s^{2}}.
\end{align}
Note that this approximation of the second derivative for sufficiently small $s$ has the leading error scaling as $O\left(s^{2}\right)$.
The operator~\eqref{eq:FiniteDifferenceLaplace} applied to the plane wave $w_{\tilde{k}}(x)$ gives
\begin{align}\label{eq:FiniteDifferenceLaplacePlaneWaves}
\begin{split}
 \frac{\partial^{2} w_{\tilde{k}}}{\partial x^{2}} \bigg|_{\tilde{\ell}} 
 & \approx N^{2} \frac{\e^{\im 2 \pi \tilde{k} \left(\tilde{\ell}+1/N\right)} - 2 \e^{\im 2 \pi \tilde{k} \tilde{\ell}} + \e^{\im 2 \pi \tilde{k} \left(\tilde{\ell}-1/N\right)}}{\sqrt{N}}
 \\
 &= -4 N^{2} \sin^{2}\left(\frac{\pi \tilde{k}}{N}\right) w_{\tilde{k}}(\tilde{\ell}).
\end{split}
\end{align}
Therefore $w_{\tilde{k}}$ is an eigenvector of the second derivative approximation~\eqref{eq:FiniteDifferenceLaplace} with eigenvalue $-4 N^{2} \sin^{2}\left(\pi \tilde{k} / N\right)$.
This enables us to identify
\begin{align}\label{eq:SecondDerivative}
 \frac{\partial^{2}}{\partial x^{2}} 
 \ \widehat{=} \,
 -4 N^{2} \QFT \sin^{2}\left(\frac{\pi \hat{k}}{N}\right) \QFT^{\dag}.
\end{align}

When we consider smooth initial conditions, we assume that the initial discretized function $|f\rangle$ has significant overlaps $|\left(\prod_{\alpha = 1}^{d} \bra{\tilde{k}_{\alpha}} \right)\ket{f}|^{2}$ only with small-wavenumber basis states $w_{\tilde{k}_{\alpha}}$ satisfying $|\tilde{k}_{\alpha}|/N \ll 1$ for all $\alpha$.
Here $\ket{\tilde{k}_{\alpha}}$ denotes the discretized version of plane wave $w_{\tilde{k}_{\alpha}}$ in dimension $\alpha$.
Then we use the small-angle approximation, $\sin\left(2 \pi \tilde{k} / N\right) \approx 2 \pi \tilde{k} / N$, to approximate the expression for the first derivative~\eqref{eq:FirstDerivative}:
\begin{align}\label{eq:FirstDerivativeSmooth}
 \frac{\partial}{\partial x} 
 \ \widehat{\approx} \
 \im 2 \pi \QFT \hat{k} \QFT^{\dag},
\end{align}
and for the second derivative~\eqref{eq:SecondDerivative}:
\begin{align}\label{eq:SecondDerivativeSmooth}
 \frac{\partial^{2}}{\partial x^{2}} 
 \ \widehat{\approx} \,
 -4 \pi^{2} \QFT \hat{k}^{2} \QFT^{\dag}.
\end{align}
When required, we characterize a smooth function by a maximum wavenumber $k_{\text{max}}$ such that the overlaps $|\left(\prod_{\alpha = 1}^{d} \bra{\tilde{k}_{\alpha}} \right)\ket{f}|^{2}$ are zero if $\tilde{k}_{\alpha} > k_{\text{max}}$ for at least one $\alpha$.
This characterization is satisfied by bandwidth-limited functions~\cite{HoLa12} but is too strict to be applicable to all smooth functions.
In general, smooth functions have decaying overlaps for increasing values of $\tilde{k}_{\alpha}$ that are non-zero for all values of $\tilde{k}_{\alpha}$.
Therefore, strictly speaking, when $k_{\text{max}}$ is used in our results, they apply to bandwidth-limited functions.
We note, however, that many smooth functions can be characterized by a certain value of $k_{\text{max}}$ such that all overlaps corresponding to $\tilde{k}_{\alpha} > k_{\text{max}}$ are sufficiently small and negligible in practical terms.

We note that, if only wave numbers in the close vicinity of a certain value $\tilde{k}^{*}$ contribute significantly to the initial conditions, then we can approximate $\sin\left(2 \pi \tilde{k} / N\right) \approx \sin\left(2 \pi \tilde{k}^{*} / N\right) + 2 \pi \cos\left(2 \pi \tilde{k}^{*} / N\right) \left(\tilde{k} - \tilde{k}^{*}\right) / N - 4 \pi^{2} \sin\left(2 \pi \tilde{k}^{*} / N\right) \left(\tilde{k} - \tilde{k}^{*}\right)^{2} / N^{2}$.
Inserting this approximation into our expressions for the first derivative~\eqref{eq:FirstDerivative} and second derivative~\eqref{eq:SecondDerivative} turns them into low-degree polynomials of $\hat{k}$ which can simplify PDE solution circuits in analogy to the small-angle approximation.

\begin{figure}
\centering
\includegraphics[width=0.95\linewidth]{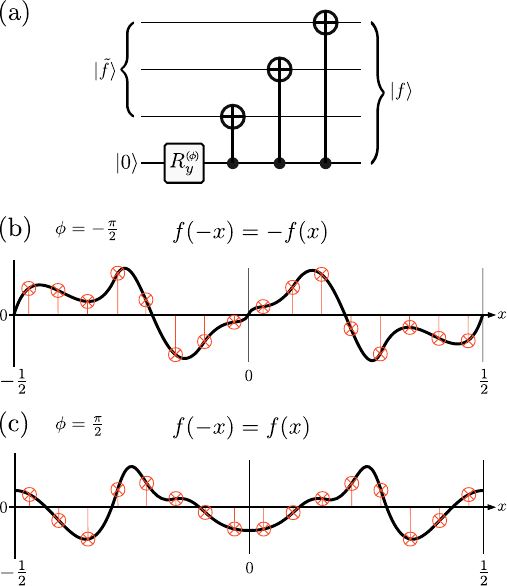}
\caption{\label{fig:BoundaryConditions}
Four-qubit quantum circuit for the realization of Dirichlet and Neumann boundary conditions using the plane wave basis of our shifted QFT.
(a) The input quantum state $\ket{\tilde{f}}$ stores the $8$ desired function values that fulfill the specific boundary conditions considered.
We add one ancilla qubit initialized in $\ket{0}$, apply $R_{y}^{(\phi)} = \e^{-\im \phi Y / 2}$ to it, where $Y$ is the Pauli $Y$ matrix, and apply the CNOTs as shown.
This circuit creates the state $\ket{f}$.
(b) To impose Dirichlet boundary conditions, we set $\phi = -\pi/2$ such that $\ket{f}$ corresponds to an odd function.
(c) To impose Neumann boundary conditions, we set $\phi = \pi/2$ such that $\ket{f}$ corresponds to an even function.
}
\end{figure}

To realize boundary conditions different from periodic ones, using the plane waves in $\mathcal{F}$ as basis functions, a procedure for Dirichlet boundary conditions $f(-1/2) = 0 = f(0)$ and Neumann boundary conditions $\partial f / \partial x\rvert_{x=-1/2} = 0 = \partial f / \partial x\rvert_{x=0}$ in one dimension is illustrated in Fig.~\ref{fig:BoundaryConditions} and explained in the figure caption.
Note that the approach requires one additional qubit and the function of interest is now defined on the interval $(-1/2, 0)$ or equivalently $(0, 1/2)$.
The key insight that makes the construction of Fig.~\ref{fig:BoundaryConditions} work is that expressing $|f\rangle$ in terms of the plane wave basis, i.e. $\ket{f} = \sum_{\tilde{k}} \braket{\tilde{k}}{f} \ket{\tilde{k}}$, for the particular symmetry choices of $\ket{f}$ in Figs.~\ref{fig:BoundaryConditions}~(b) and (c), leads to an expansion of $\ket{f}$ in terms of sine (b) and cosine (c) functions, which are appropriate bases for the specific instances of boundary conditions addressed in Fig.~\ref{fig:BoundaryConditions}.
Generalizing the strategy to $d$ dimensions is straightforward since the necessary reflections are independently implemented in the $d$ dimensions using $d$ additional qubits.

In the paper, we use variants of QSVT~\cite{LoYoCh16, LoCh17, GiEtAl19, LoCh19, PeEtAl21, MaEtAl21, LiBoAo22, MaEtAl23, SiEtAl23, MoWi24} (see~\cite{DoWhLi22, KiEtAl23, DeEtAl23, KaEtAl25} for experimental realizations) to design quantum circuits that represent Fourier series and polynomials of operators.

The Fourier-based QSVT construction for the Hermitian operator $\mathcal{H}$ of interest requires one ancilla qubit and the unitary operator $U_{\mathcal{H}} = |0\rangle_{\text{anc}} \langle 0|_{\text{anc}} \otimes \mathds{1} + |1\rangle_{\text{anc}} \langle 1|_{\text{anc}} \otimes \e^{\im \mathcal{H}}$.
The desired Fourier series $\sum_{\zeta = -D}^{D} c_{\zeta} \e^{\im \zeta \mathcal{H}}$ for a positive integer $D$ is realized by a quantum circuit via, firstly, initialization of the ancilla qubit in the state $\ket{0}_{\text{anc}}$, secondly, $2D+1$ repetitions of $U_{\mathcal{H}}$ each surrounded by certain single-ancilla-qubit unitaries that can be determined efficiently given the Fourier coefficients $c_{\zeta}$ and, thirdly, post-selection of the ancilla qubit being measured in $\ket{0}_{\text{anc}}$.

To prepare the polynomial-based QSVT circuits, we block-encode the operator $\mathcal{H}$ of interest inside a larger unitary operator $V_{\mathcal{H}}$ using ancilla qubits such that $\mathcal{H} = \bra{\boldsymbol{0}}_{\text{anc}} V_{\mathcal{H}} \ket{\boldsymbol{0}}_{\text{anc}}$ where $\ket{\boldsymbol{0}}_{\text{anc}} = \ket{0, 0, \ldots, 0}_{\text{anc}}$ is the initial state of the ancilla qubits.
We create the block-encoding using an approach based on a linear combination of unitary operators (LCU)~\cite{ChWi12}.
Then the polynomial-based QSVT approach defines the circuit $\prod_{\zeta = 1}^{D} \left[W_{\text{anc}}(\phi_{\zeta}) V_{\mathcal{H}}\right]$ that contains the ancilla-qubit gates $W_{\text{anc}}\left(\phi\right)$ where the $D$ angles $\phi_{\zeta}$ are chosen in such a way that $\bra{\boldsymbol{0}}_{\text{anc}} \prod_{\zeta = 1}^{D} \left[W_{\text{anc}}(\phi_{\zeta}) V_{\mathcal{H}}\right] \ket{\boldsymbol{0}}_{\text{anc}}$ creates the desired degree-$D$ polynomial of $\mathcal{H}$.
Throughout the paper, the encountered operators $\mathcal{H}$ are Hermitian and we can understand a polynomial of an operator $\mathcal{H}$ as a polynomial of the eigenvalues of $\mathcal{H}$ \footnote{Therefore, strictly speaking, we work with a special case of QSVT referred to as quantum eigenvalue transformation~\cite{MaEtAl21}.}.
Importantly, as a function of the eigenvalues of $\mathcal{H}$, a polynomial-based QSVT circuit of $D$ layers creates either an odd function, if $D$ is odd, or an even function, if $D$ is even.
To handle functions that are neither odd nor even using the polynomial-based QSVT strategy, we shift the entire spectrum of eigenvalues of $\mathcal{H}$ to the non-negative real numbers and then apply the QSVT formalism to the shifted $\mathcal{H}$.
We provide the block-encodings of the shifted operators $\mathcal{H}$ as well as their largest and smallest eigenvalues.
This knowledge about the eigenvalue spectrum can be used when the angles $\phi_{\zeta}$ are numerically computed by software libraries such as pyQSP~\cite{GiEtAl19, Ha19, ChEtAl20, DoEtAl21, MaEtAl21} and can improve the accuracy of the polynomial approximations.

We also utilize Trotter product formulas~\cite{HaSu05} to approximately decompose exponentials of operators into quantum gates.
More specifically, we present quantum circuits resulting from first-order Trotter formulas.
Circuits corresponding to higher-order Trotter formulas can readily be derived from the presented results.

\section{Incompressible advection equation}
\label{sec:AdvectionEquation}

We consider the advection equation for incompressible flows in $d$ spatial dimensions:
\begin{align}\label{eq:AdvectionEquation}
 \frac{\partial}{\partial t} f 
 = -\left(r_{1} \frac{\partial}{\partial x_{1}} + r_{2} \frac{\partial}{\partial x_{2}} + \ldots + r_{d} \frac{\partial}{\partial x_{d}}\right) f
\end{align}
where $f = f(x_{1}, x_{2}, \ldots, x_{d}, t)$ and $r_{1}$, $r_{2}$, \ldots, $r_{d}$ are real numbers.
Using Eq.~\eqref{eq:FirstDerivative} we express the discretized Eq.~\eqref{eq:AdvectionEquation} as
\begin{align}\label{eq:AdvectionSchroedingerEquation}
 \frac{\partial}{\partial t} \ket{f}
 = -\im N \sum_{\alpha=1}^{d} r_{\alpha} \QFT_{\alpha} \sin\left(\frac{2 \pi \hat{k}_{\alpha}}{N}\right) \QFT_{\alpha}^{\dag} \ket{f}.
\end{align}
The solution to Eq.~\eqref{eq:AdvectionSchroedingerEquation} is given by
\begin{align}\label{eq:TimeEvolutionAdvection}
 \ket{f(t)} 
 = \QFT^{\otimes d} \prod_{\alpha=1}^{d} \e^{-\im t N r_{\alpha} \sin\left(2 \pi \hat{k}_{\alpha} / N\right)} \left(\QFT^{\dag}\right)^{\otimes d} \ket{f(0)}
\end{align}
where the time evolution operator is a product of $d$ independent one-dimensional operators.
In our notation $\ket{f(0)}$ represents the initial conditions, i.e.\ the discretized $f(x_{1}, x_{2}, \ldots, x_{d}, t=0)$.

\begin{figure}
\centering
\includegraphics[width=0.8\linewidth]{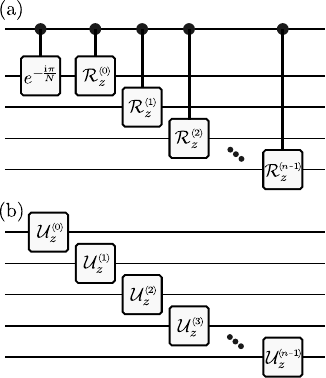}
\caption{\label{fig:Advection}
Quantum circuits for the one-dimensional incompressible advection equation.
(a) Circuit representation of the operator $U_{\hat{k}}$~\eqref{eq:PlaneWaveBlockEncoding} where $\mathcal{R}_{z}^{(\zeta)} = \e^{-\im \pi 2^{-\zeta - 1} Z_{\zeta}}$.
(b) Circuit representation of~\eqref{eq:SmoothAdvection} where $\mathcal{U}_{z}^{(\zeta)} = \e^{\im \pi t r 2^{-\zeta+n-1} Z_{\zeta}}$ and we exclude the global phase $\e^{\im \pi t r}$.
}
\end{figure}

In the following we focus on the one-dimensional case and omit the index $\alpha$ for clarity.
There exist several ways in which we can represent the one-dimensional time evolution operator for Eq.~\eqref{eq:TimeEvolutionAdvection} in terms of an explicit quantum circuit.
One possibility is to use the truncated Jacobi-Anger expansion
\begin{align}\label{eq:JacobiAnger}
 \e^{-\im t N r \sin\left(2 \pi \hat{k} / N\right)}
 \approx
 \sum_{\zeta = -D}^{D} J_{\zeta}(-t N r) \e^{\im 2 \pi \hat{k} \zeta / N}
\end{align}
where $J_{\zeta}$ is a Bessel function of the first kind and, for error $\epsilon$, $D$ needs to scale as $\Theta\left(t N r + \frac{\log(1/\epsilon)}{\log\left[\e + \log(1/\epsilon) / (t N r)\right]}\right)$ (see~\cite{LoCh17, LoCh19} and Corollary 32 in~\cite{GiEtAl19}).
The right-hand side of Eq.~\eqref{eq:JacobiAnger} can be realized using Fourier-based QSVT (see Sec.~\ref{sec:Background}) with the operator
\begin{align}\label{eq:PlaneWaveBlockEncoding}
 U_{\hat{k}} 
 = 
 |0\rangle_{\text{anc}} \langle 0|_{\text{anc}} \otimes \mathds{1} + |1\rangle_{\text{anc}} \langle 1|_{\text{anc}} \otimes \e^{\im 2 \pi \hat{k} / N},
\end{align}
where
\begin{align}
 \e^{\im 2 \pi \hat{k} / N} = \e^{-\im \pi / N} \prod_{\beta = 0}^{n-1} \e^{-\im \pi 2^{-\beta-1} Z_{\beta}},
\end{align}
using the definition of $\hat{k}$ in Eq.~\eqref{eq:HatK}.
The quantum circuit representation of Eq.~\eqref{eq:PlaneWaveBlockEncoding} is shown in Fig.~\ref{fig:Advection}~(a).

Another possibility to construct a circuit for Eq.~\eqref{eq:TimeEvolutionAdvection} uses the discrete Fourier transform to express
\begin{align}\label{eq:AdvectionEquationDFT}
 \e^{-\im t N r \sin\left(2 \pi \hat{k} / N\right)} 
 = \sum_{\zeta = -N/2}^{N/2 - 1} c_{\zeta} \e^{\im 2 \pi \zeta \hat{k} / N}
\end{align}
where
\begin{align}
 c_{\zeta} = \frac{1}{N} \sum_{\tilde{k} = -N/2}^{N/2 - 1} \e^{-\im t N r \sin\left(2 \pi \tilde{k} / N\right)} \e^{-\im 2 \pi \tilde{k} \zeta / N}.
\end{align}
A circuit of depth scaling as $O(N)$ for the operator of Eq.~\eqref{eq:AdvectionEquationDFT} can be created using standard Fourier-based QSVT with $D = N/2$, $c_{N/2} = 0$ and Eq.~\eqref{eq:PlaneWaveBlockEncoding}.

Summarizing the two preceding circuit constructions and taking into account the circuit depth $n$ of the operator~\eqref{eq:PlaneWaveBlockEncoding}, we conclude that the time evolution operator in Eq.~\eqref{eq:TimeEvolutionAdvection}, excluding the QFT, has a circuit representation of depth scaling as $O\left(n \min[N, t N r_{\text{max}} + \log(1/\epsilon)]\right)$ where $r_{\text{max}}$ is the $r_{\alpha}$ of largest magnitude.
Here we use $d$ ancilla qubits to create $d$ independent one-dimensional time evolution operators in parallel for the realization of~\eqref{eq:PlaneWaveBlockEncoding}.
We note that, if a sufficiently small $\epsilon$ is used in the Fourier-based QSVT procedure, the probability of success is close to $1$, i.e.\ $\langle f(t) | f(t) \rangle / \langle f(0) | f(0) \rangle \lesssim 1$, because the time evolution operator in Eq.~\eqref{eq:TimeEvolutionAdvection} is unitary.

We can reduce the required number of ancilla qubits by realizing~\eqref{eq:TimeEvolutionAdvection} sequentially.
To achieve this, we apply the one-dimensional time evolution operator to the different dimensions one after another, each time reusing the same ancilla qubit initialized and measured in $|0\rangle_{\text{anc}}$.
The circuit depth of this approach scales as $O\left(d n \min[N, t N r_{\text{max}} + \log(1/\epsilon)]\right)$ and the construction needs only one ancilla qubit.

We can derive more efficient circuits for smooth initial states $|f(0)\rangle$.
With the help of Eq.~\eqref{eq:FirstDerivativeSmooth} we obtain
\begin{eqnarray}\label{eq:SmoothAdvection}
 \e^{-\im t N r \sin\left(2 \pi \hat{k} / N\right)} 
 & \approx & \e^{-\im 2 \pi t r \hat{k}}
 \nonumber\\
 & = & \e^{\im \pi t r} \prod_{\beta=0}^{n-1} \e^{\im \pi t r 2^{-\beta+n-1} Z_{\beta}}
\end{eqnarray}
which is implemented with the particularly simple circuit, of $O(1)$ depth, shown in Fig.~\ref{fig:Advection}~(b).

\section{Heat equation}
\label{sec:HeatEquation}

Let us now examine the heat equation in $d$ spatial dimensions:
\begin{align}\label{eq:HeatEquation}
 \frac{\partial}{\partial t} f 
 =
 u \left( \frac{\partial^{2}}{\partial x_{1}^{2}} + \frac{\partial^{2}}{\partial x_{2}^{2}} + \ldots + \frac{\partial^{2}}{\partial x_{d}^{2}} \right) f
\end{align}
where $f = f(x_{1}, x_{2}, \ldots, x_{d}, t)$ and $u$ is a positive real number.
Via Eq.~\eqref{eq:SecondDerivative} we write the discretized Eq.~\eqref{eq:HeatEquation} as
\begin{align}\label{eq:HeatEquationDiscretized}
 \frac{\partial}{\partial t} \ket{f} 
 =
 -4 N^{2} u \sum_{\alpha=1}^{d} \QFT_{\alpha} \sin^{2}\left(\frac{\pi \hat{k}_{\alpha}}{N}\right) \QFT_{\alpha}^{\dag} |f\rangle.
\end{align}
Equation~\eqref{eq:HeatEquationDiscretized} has the solution
\begin{align}\label{eq:TimeEvolutionHeat}
 \ket{f(t)}
 = 
 \QFT^{\otimes d} \prod_{\alpha=1}^{d} \e^{-t 4 N^{2} u \sin^{2}\left(\pi \hat{k}_{\alpha} / N\right)} \left(\QFT^{\dag}\right)^{\otimes d} \ket{f(0)},
\end{align}
where the time evolution operator is a product of $d$ independent one-dimensional time evolution operators.

In the following, we describe several quantum circuit designs for the time evolution operator in~\eqref{eq:TimeEvolutionHeat} for which it is sufficient to consider the one-dimensional case, and we henceforth omit the index $\alpha$.

One approach for the circuit construction adapts the procedure~\cite{ChSo17} to our problem.
To that end we approximate the Fourier transform of the Gaussian,
\begin{align}\label{eq:ApproximateGaussian}
\begin{split}
 &\e^{-t 4 N^{2} u \sin^{2}\left(\pi \tilde{k} / N\right)} 
 \\
 &= \frac{1}{\sqrt{2 \pi}} \int_{-\infty}^{\infty} \diff\omega\, 
 \e^{-\omega^{2} / 16} \e^{-\im \sqrt{t u} \omega N \sin\left(\pi \tilde{k} / N\right)}
 \\
 &\approx \frac{1}{\sqrt{2 \pi}} \sum_{\zeta = -G}^{G}\delta_{\omega} \e^{-(\zeta \delta_{\omega})^{2} / 16} \e^{-\im \sqrt{t u} \zeta \delta_{\omega} N \sin\left(\pi \tilde{k} / N\right)},
\end{split}
\end{align}
where, for error $\epsilon$ and provided the assumptions of~\cite{ChSo17} are satisfied, $\delta_{\omega}$ needs to scale as $\Theta\left(\left[N \sqrt{t u \log(1/\epsilon)} \right]^{-1}\right)$ and $G$ as $O\left[ N \sqrt{t u} \log(1/\epsilon) \right]$.
We define $C_{0} = \delta_{\omega} / \sqrt{2\pi}$, $C_{1} = \delta_{\omega}^{2} / 16$ and $C_{2} = N\sqrt{t u} \delta_{\omega}$, such that Eq.~\eqref{eq:ApproximateGaussian} now reads
\begin{align}\label{eq:ApproximateGaussian2}
 C_{0} \sum_{\zeta = -G}^{G} \e^{-C_{1} \zeta^{2}} \e^{-\im C_{2} \zeta \sin\left(\pi \tilde{k} / N\right)}.
\end{align}
Following~\cite{ChSo17}, for $\e^{-\im C_{2} \zeta \sin\left(\pi \tilde{k} / N\right)}$ we use an approximation of error $O(\epsilon)$, which here we obtain via the truncated Jacobi-Anger expansion,
\begin{align}\label{eq:truncated_JAexpansion}
 \e^{-\im C_{2} \zeta \sin\left(\pi \tilde{k} / N\right)} 
 \approx 
 \sum_{\eta = -D}^{D} J_{\eta}(-C_{2} \zeta) \e^{\im \pi \tilde{k} \eta / N},
\end{align}
where, for error $\epsilon$, $D$ scales like
\begin{align}
\begin{split}
 &\Theta\left(
 C_{2}G + \frac{\log(1/\epsilon)}{\log\left[\e + \log(1/\epsilon)/(C_{2}G)\right]}
 \right) 
 \\
 &= O\left[C_{2} G + \log(1/\epsilon)\right]
 \\
 &= O\left[\sqrt{t u} N \log(1/\epsilon)\right].
\end{split}
\end{align}
With help from the truncated Jacobi-Anger series, we approximate Eq.~\eqref{eq:ApproximateGaussian2} by
\begin{align}
 \sum_{\eta = -D}^{D} \left[C_{0} \sum_{\zeta = -G}^{G} \e^{-C_{1} \zeta^{2}} J_{\eta}(-C_{2} \zeta) \right] \e^{\im \pi \tilde{k} \eta / N}.
\end{align}
Note that the error $\epsilon$ in Eq.~\eqref{eq:truncated_JAexpansion} leads to an error of $O(\epsilon)$ in \eqref{eq:ApproximateGaussian2},
\begin{align}
\begin{split}
 &\bigg|
 \frac{1}{\sqrt{2 \pi}} \sum_{\zeta = -G}^{G}\delta_{\omega} \e^{-(\zeta \delta_{\omega})^{2} / 16} \epsilon
 \bigg|
 \\
 &\le
 \epsilon+
 \epsilon \bigg|
 \frac{1}{\sqrt{2 \pi}} \int_{-\infty}^{\infty}\diff\omega\, \e^{-\omega^{2} / 16}
 \bigg|
 \\
 &=O(\epsilon).
\end{split}
\end{align}
We conclude with the final expression
\begin{align}\label{eq:HeatEquation1DFourier}
 \e^{-t 4 N^{2} u \sin^{2}\left(\pi \hat{k} / N\right)} 
 \approx \sum_{\eta = -D}^{D} c_{\eta} \e^{\im \pi \hat{k} \eta / N},
\end{align}
where
\begin{align}
 c_{\eta} = C_{0} \sum_{\zeta = -G}^{G} \e^{-C_{1} \zeta^{2}} J_{\eta}(-C_{2} \zeta).
\end{align}
The finite Fourier series on the right-hand side of Eq.~\eqref{eq:HeatEquation1DFourier} can be realized using Fourier-based QSVT with the operator
\begin{align}\label{eq:PlaneWaveBlockEncodingHeat}
 U_{\hat{k}} = |0\rangle_{\text{anc}} \langle 0|_{\text{anc}} \otimes \mathds{1} + |1\rangle_{\text{anc}} \langle 1|_{\text{anc}} \otimes \e^{\im \pi \hat{k} / N}
\end{align}
where
\begin{align}
 \e^{\im \pi \hat{k} / N} = \e^{-\im \pi / (2 N)} \prod_{\beta = 0}^{n-1} \e^{-\im \pi 2^{-\beta-2} Z_{\beta}}.
\end{align}
The quantum circuit for the operator of Eq.~\eqref{eq:PlaneWaveBlockEncodingHeat} is equivalent to the one shown in Fig.~\ref{fig:Advection} (a) except that the angle of each gate in Fig.~\ref{fig:Advection} (a) needs to be divided by $2$.

Alternatively we can construct a quantum circuit utilizing the discrete Fourier transform to write
\begin{align}\label{eq:HeatEquationDFT}
 \e^{-t 4 N^{2} u \sin^{2}\left(\pi \hat{k} / N\right)} 
 = \sum_{\zeta = -N/2}^{N/2 - 1} c_{\zeta} \e^{\im 2 \pi \zeta \hat{k} / N}
\end{align}
where
\begin{align}
 c_{\zeta} = \frac{1}{N} \sum_{\tilde{k} 
 = -N/2}^{N/2 - 1} \e^{-t 4 N^{2} u \sin^{2}\left(\pi \tilde{k} / N\right)} \e^{-\im 2 \pi \tilde{k} \zeta / N}.
\end{align}
Then Fourier-based QSVT with $D = N/2$, $c_{N/2} = 0$ and the operator of Eq.~\eqref{eq:PlaneWaveBlockEncoding} enables us to create a circuit of depth $O(N)$ representing Eq.~\eqref{eq:HeatEquationDFT}.

We conclude that the time evolution operator corresponding to the one-dimensional heat equation can be prepared by a quantum circuit of depth scaling as $O\left(n N \min\left[1, \sqrt{t u} \log(1/\epsilon)\right]\right)$, excluding the known depth of the QFT.
For the $d$-dimensional heat equation, we run $d$ copies of the quantum circuit for the one-dimensional heat equation either in parallel using $d$ ancilla qubits or sequentially by reusing one ancilla qubit.

We emphasize that, assuming $\epsilon$ is sufficiently small, the success probability of the Fourier-based QSVT approach is approximately
\begin{align}
 p = \frac{\langle f(t)| f(t)\rangle}{\langle f(0)| f(0)\rangle}
\end{align}
and satisfies $\exp(-t 8 N^{2} u d) \leq p \leq 1$ which readily follows from Eq.~\eqref{eq:TimeEvolutionHeat} whereby the lower (upper) bound for $p$ is obtained if $|f(0)\rangle$ is a product state of discretized plane waves of wavenumber $\tilde{k} = -N/2$ $\left(\tilde{k} = 0\right)$.
For smooth $|f(0)\rangle$, the success probability fulfills $\exp(-t 8 \pi^{2} k_{\text{max}}^{2} u d) \leq p \leq 1$.
To achieve the successful preparation of the state $|f(t)\rangle$ in Eq.~\eqref{eq:TimeEvolutionHeat}, the quantum circuit needs to be rerun a number of times that is on average approximately $1/p$, which can be a large number.

For smooth initial functions $\ket{f(0)}$ we can use the approximation~\eqref{eq:SecondDerivativeSmooth} such that
\begin{align}\label{eq:SmoothHeat}
\begin{split}
 &\e^{-t 4 N^{2} u \sin^{2}\left(\pi \hat{k} / N\right)} 
 \approx \e^{-4 \pi^{2} t u \hat{k}^{2}}
 \\
 &\quad
 = \e^{-\pi^{2} t u \left(N^{2}+2\right)/3} \prod_{\beta=0}^{n-1} \e^{-\pi^{2} t u N 2^{-\beta} Z_{\beta}}
 \\
 &\qquad\quad
 \times \prod_{\zeta = 0}^{n-1} \prod_{\eta = \zeta+1}^{n-1} \e^{-\pi^{2} t u N^{2} 2^{-\zeta-\eta-1} Z_{\zeta} Z_{\eta}},
\end{split}
\end{align}
where we use the definition of $\hat{k}$ in Eq.~\eqref{eq:HatK}.

\begin{figure}
\centering
\includegraphics[width=\linewidth]{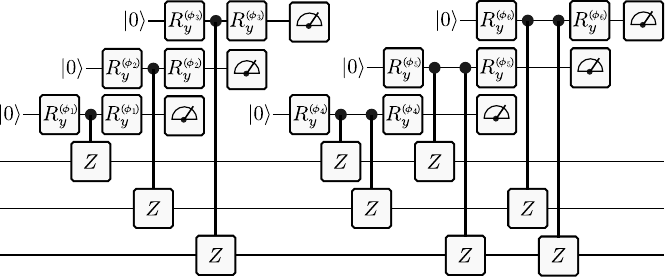}
\caption{\label{fig:Heat}
Quantum circuit realization of the operator product in Eq.~\eqref{eq:SmoothHeat} for system size $n = 3$ using three ancilla qubits that are assumed to be measured in $\ket{0}$ and then reused.
The angles $\phi_{1}$, $\phi_{2}$, \ldots, $\phi_{6}$ are calculated via Eq.~\eqref{eq:ThetaToPhi} using $\theta_{1} = -8 \pi^{2} t u$, $\theta_{2} = -4 \pi^{2} t u$, $\theta_{3} = -2 \pi^{2} t u$, $\theta_{4} = -16 \pi^{2} t u$, $\theta_{5} = -8 \pi^{2} t u$, $\theta_{6} = -4 \pi^{2} t u$, respectively.
Note that one can readily reduce the required number of ancilla qubits from three to one by redefining the circuit such that only one ancilla qubit is sequentially measured and reused.
}
\end{figure}

To realize the non-unitary operators of Eq.~\eqref{eq:SmoothHeat}, which are exponentials of Pauli strings, we note that any Pauli string $P$ satisfies $P^{2} = \mathds{1}$ such that
\begin{align}\label{eq:ExpPauliString}
 \e^{\theta P} = \cosh(\theta) \mathds{1} + \sinh(\theta) P.
\end{align}
The right-hand side of Eq.~\eqref{eq:ExpPauliString} is a weighted sum of two unitary operators, $\mathds{1}$ and $P$.
Therefore Eq.~\eqref{eq:ExpPauliString} can be represented, up to a normalization factor, by a block-encoding quantum circuit using one ancilla qubit whose two computational basis states $\ket{0}$ and $\ket{1}$ control the application of $\mathds{1}$ and $P$, respectively.
The corresponding quantum gate is a controlled-$P$ gate that is equivalent to the product of the controlled single-qubit-Pauli-matrix gates for all Pauli matrices in $P$.
To create the desired block-encoding, we initialize the ancilla qubit in $\ket{0}$, apply $R_{y}^{(\phi)}$ to it with the angle
\begin{align}\label{eq:ThetaToPhi}
 \phi = 
 2 \arccos\left(\sqrt{\frac{\cosh(\theta)}{\cosh(\theta)+\sinh(|\theta|)}}\right),
\end{align}
then insert the controlled-$P$ gate, and finally apply $R_{y}^{\left(-\text{sign}(\theta)\phi\right)}$ to the ancilla qubit.
Now measuring the ancilla qubit in the state $\ket{0}$ creates the operator
\begin{align}\label{eq:NormalizedPauliExp}
 \frac{\cosh(\theta) \mathds{1} + \sinh(\theta) P}{\cosh(\theta)+\sinh(|\theta|)} = \frac{\e^{\theta P}}{\e^{|\theta|}},
\end{align}
where we use $\cosh(\theta) = \cosh(|\theta|)$ and $\sinh(\theta) = \text{sign}(\theta) \sinh(|\theta|)$.
For the realization of a product of several non-unitary operators of the form~\eqref{eq:NormalizedPauliExp}, Fig.~\ref{fig:Heat} shows an example circuit. 

If our goal is to reduce the circuit depth for the preparation of~\eqref{eq:SmoothHeat}, we apply in parallel as many two-qubit operators of~\eqref{eq:SmoothHeat} as possible.
To maximize parallelization, we apply the operators in groups, one group after another, where each group contains all two-qubit operators that have the same distance between the two qubits they act upon.
It is straightforward to see that every group of gates can be realized in $2$ consecutive steps where in each step $\approx n/2$ non-overlapping operators of~\eqref{eq:SmoothHeat} are applied in parallel.
This approach leads to $O(n)$ circuit depth and requires $O(n)$ ancilla qubits.

If our goal is to reduce the number of ancilla qubits, we create the product of non-unitary operators in Eq.~\eqref{eq:SmoothHeat} sequentially using one ancilla qubit that probabilistically, upon successive measurements of it being in the state $\ket{0}$, realizes each term in Eq.~\eqref{eq:SmoothHeat} one after another.

The overall probability $p$ of successfully creating the state proportional to $\ket{f(t)}=\e^{\sum_{\beta = 1}^{K} \theta_{\beta} P_{\beta}}\ket{f(0)}$ is
\begin{align}
\begin{split}
 p
 &= \e^{-2 \sum_{\beta = 1}^{K} |\theta_{\beta}|}
 \frac{\braket{f(t)}{f(t)}}{\braket{f(0)}{f(0)}}
 \\
 &= \e^{-4 \pi^{2} t u \left(N^{2}-1\right)/3}
 \frac{\braket{f(t)}{f(t)}}{\braket{f(0)}{f(0)}}.
\end{split}
\end{align}
Here we denote the total number of operators in Eq.~\eqref{eq:SmoothHeat} by $K$, which scales like $O(n^{2})$, evaluate $\sum_{\beta} |\theta_{\beta}|$ for the specific $\theta_{\beta}$ of Eq.~\eqref{eq:SmoothHeat} and ignore the first factor in Eq.~\eqref{eq:SmoothHeat} because it is just a scalar.

We readily generalize the procedure to the $d$-dimensional scenario by implementing the circuits corresponding to the different dimensions independently and in the same way as was just described for the one-dimensional heat equation.
Then the overall success probability is
\begin{align}\label{eq:HeatSuccessProbability}
 p = 
 \exp\left[
    -\frac{4}{3} \pi^{2} t \left(N^{2}-1\right) d u
\right] \frac{\braket{f(t)}{f(t)}}{\braket{f(0)}{f(0)}}.
\end{align}
The circuit depth scales as either $O(n)$ using $O(dn)$ ancilla qubits, $O(n^{2})$ using $d$ ancilla qubits, or $O(d n^{2})$ using one ancilla qubit whereby the circuits for the $d$ dimensions are sequentially run.
We emphasize that, however, for the successful realization of the desired state, the circuit needs to be rerun on average $1/p$ times, which can be a large number.
In comparison with the Fourier-based QSVT approach, which has a success probability of $\braket{f(t)}{f(t)}/\braket{f(0)}{f(0)}$ and requires an advanced circuit architecture, here we have the additional factor of $\exp\left[-\frac{4}{3} \pi^{2} t \left(N^{2}-1\right) d u\right]$ in the success probability but we also have a significantly simpler circuit structure.

To avoid the additional factor of $\exp\left[-\frac{4}{3} \pi^{2} t \left(N^{2}-1\right) d u\right]$ in the success probability, we can construct a quantum circuit for $\exp\left(-4 \pi^{2} t u \hat{k}^{2}\right)$ by means of the approach~\cite{ChSo17} used above.
Now this approach leads to the operator approximation
\begin{align}\label{eq:ApproximateGaussianSmooth}
\begin{split}
 &\e^{-4 \pi^{2} t u \hat{k}^{2}} 
 \\
 &\approx \frac{1}{\sqrt{2 \pi}} \sum_{\zeta = -G}^{G} \delta_{\omega} \e^{-(\zeta \delta_{\omega})^{2} / 16} \e^{-\im \pi \sqrt{t u} \zeta \delta_{\omega} \hat{k}}
 \\
 &= \sum_{\zeta = -G}^{G} c_{\zeta} \e^{-\im \pi C_{3} \hat{k} \zeta}
\end{split}
\end{align}
where $c_{\zeta} = \exp\left[-(\zeta \delta_{\omega})^{2} / 16\right] / \sqrt{2 \pi}$ and $C_{3} = \sqrt{t u} \delta_{\omega}$.
To achieve error $\epsilon$ under the assumptions of~\cite{ChSo17}, $\delta_{\omega}$ needs to scale as $\Theta\left(\left[\sqrt{t u \log(1/\epsilon)}\right]^{-1}\right)$ and $G$ as $O\left[\sqrt{t u} \log(1/\epsilon)\right]$. 
The final expression in Eq.~\eqref{eq:ApproximateGaussianSmooth} can be turned into a quantum circuit through Fourier-based QSVT with the operator
\begin{align}
 U_{\hat{k}} = |0\rangle_{\text{anc}} \langle 0|_{\text{anc}} \otimes \mathds{1} + |1\rangle_{\text{anc}} \langle 1|_{\text{anc}} \otimes \e^{-\im \pi C_{3} \hat{k}}
\end{align}
which results in the total circuit depth scaling as $O\left[n \sqrt{t u} \log(1/\epsilon)\right]$.

\section{Isotropic acoustic wave equation}
\label{sec:WaveEquation}

In the following we study the isotropic acoustic wave equation in $d$ spatial dimensions:
\begin{align}\label{eq:WaveEquation}
 \frac{\partial^{2}}{\partial{t}^{2}} f = v^{2} \left(\frac{\partial^{2}}{\partial{x}^{2}_{1}} + \frac{\partial^{2}}{\partial{x}^{2}_{2}} + \ldots + \frac{\partial^{2}}{\partial{x}^{2}_{d}}\right)f
\end{align}
where $f = f(x_{1}, x_{2}, \ldots, x_{d}, t)$ and $v$ is real-valued.
We denote the spatially discretized form of $f$ as $\ket{f}$ and its first- and second-order (continuous) time derivatives as $\frac{\partial}{\partial t}\ket{f} =\partial_{t}\ket{f}=\ket{\partial_{t}f}$ and $\frac{\partial^{2}}{\partial t^{2}}\ket{f} =\partial^{2}_{t}\ket{f}=\ket{\partial_{t}^{2}f}$, respectively.
The operator $\Delta_{\alpha}$ represents the discretized Laplace operator acting on dimension $\alpha$ and we denote by $\Delta^{(d)}$ the sum of the discretized Laplacians over all $d$ spatial dimensions, $\Delta^{(d)}=\sum_{\alpha=1}^{d}\Delta_{\alpha}$.
We also define
\begin{align}\label{eq:SqrtMinusLaplace}
 \mathcal{O}_{\alpha} 
 = \sqrt{-\Delta_{\alpha}}
 = \QFT_{\alpha} \mathcal{D}_{\alpha} \QFT_{\alpha}^{\dag}
\end{align}
where $\mathcal{D}_{\alpha}$ is a diagonal matrix with entries that are the square root of minus the eigenvalues of the Laplace operator in dimension $\alpha$.
Note that $\mathcal{O}_{\alpha}$ is Hermitian,
\begin{align}\label{eq:OIsHermitian}
 \mathcal{O}_{\alpha}^{\dag} = \mathcal{O}_{\alpha},
\end{align}
because the Laplace operator is Hermitian.
We use Eq.~\eqref{eq:SecondDerivative} to obtain
\begin{align}\label{eq:EigenvaluesSqrtMinusLaplace}
 \mathcal{D}_{\alpha} 
 = 2 N \sin\left(\frac{\pi \hat{k}_{\alpha}}{N}\right).
\end{align}

To construct circuits to solve the wave equation, we first consider $K$ decoupled wave equations
\begin{align}\label{eq:SubspaceWaveEquation}
 \ket{\zeta} \otimes \ket{\partial_{t}^{2} f} = \left( v^{2} \mathds{1} \otimes \Delta^{(d)} \right) \left( \ket{\zeta} \otimes \ket{f} \right)
\end{align}
where $\ket{\zeta}$ is one of the basis vectors $\{\ket{0}, \ket{1}, \ldots, \ket{K-1}\}$ that can be used to represent the identity operator
\begin{align}
 \mathds{1} = \sum_{\zeta=0}^{K-1} |\zeta\rangle \langle\zeta|.
\end{align}
We denote the square root of the block-diagonal matrix of Laplacians multiplied by $-v^{2}$ by
\begin{align}\label{eq:SqrtDiagMatLap}
    \mathcal{H}^{(d)} = \sqrt{-v^{2}\mathds{1}\otimes\Delta^{(d)}},
\end{align}
and note that solutions to the wave equations Eq.~\eqref{eq:SubspaceWaveEquation} must also satisfy
\begin{align}\label{eq:SubspaceWaveEquationA}
\begin{split}
 &\ket{\zeta} \otimes \ket{\partial_{t}^{2} f} -\text{i} \mathcal{H}^{(d)} \left( \ket{\zeta} \otimes \ket{\partial_{t}f} \right) 
 \\ 
 &= \left( v^{2} \mathds{1} \otimes \Delta^{(d)} \right) \left( \ket{\zeta} \otimes \ket{f} \right) -\text{i} \mathcal{H}^{(d)} \left( \ket{\zeta} \otimes \ket{\partial_{t}f} \right),
\end{split}
\end{align}
where we have simply added $-\text{i}\mathcal{H}^{(d)} \left( \ket{\zeta} \otimes \ket{\partial_{t}f} \right)$
to both sides of Eq.~\eqref{eq:SubspaceWaveEquation}.
Using Eq.~\eqref{eq:SqrtDiagMatLap} we can rewrite Eq.~\eqref{eq:SubspaceWaveEquation} as the Schr\"{o}dinger equation
\begin{align}\label{eq:WaveSchrodinger}
 \partial_{t}\ket{\psi}=-\text{i}\mathcal{H}^{(d)}\ket{\psi},
\end{align}
where we have defined the state
\begin{align}\label{eq:WaveSchrodingerStateA}
 \ket{\psi} \propto \ket{\zeta}\otimes\ket{\partial_{t}f}  -\text{i}\mathcal{H}^{(d)}\ket{\zeta}\otimes \ket{f},
\end{align}
up to a normalizing constant.
We therefore interpret $\mathcal{H}^{(d)}$ as a time-independent Hamiltonian such that the formal solution of Eq.~\eqref{eq:WaveSchrodinger} is
\begin{align}\label{eq:WaveUnitary}
 \ket{\psi(t)} = \exp\left(-\text{i}t\mathcal{H}^{(d)}\right)\ket{\psi(0)}.
\end{align}

To construct a representation of the Hamiltonian $\mathcal{H}^{(d)}$ that is appropriate for the implementation of the unitary evolution Eq.~\eqref{eq:WaveUnitary} on a gate-based quantum computer, we express $\mathcal{H}^{(d)}$ as the sum
\begin{align}\label{eq:WaveHamiltonianAnsatz}
 \mathcal{H}^{(d)}= v\sum_{\alpha=1}^{d}\gamma_{\alpha}\otimes\mathcal{O}_{\alpha},
\end{align}
where $\gamma_{\alpha}$ are operators of the form
\begin{align}\label{eq:GammaBasis}
 \gamma_{\alpha}=\sum_{\nu,\eta=0}^{K-1}\left(\gamma_{\alpha}\right)_{\nu,\eta}\ket{\nu}\bra{\eta}.
\end{align}
We then seek a set of Hermitian operators $\{\gamma_{\alpha}\}$ such that
\begin{align}\label{eq:WaveHamiltonianAnsatzRequired}
 \left[\mathcal{H}^{(d)}\right]^{2}=-v^{2}\mathds{1}\otimes \Delta^{(d)},
\end{align}
in accordance with Eq.~\eqref{eq:SqrtDiagMatLap}.
By squaring the sum Eq~\eqref{eq:WaveHamiltonianAnsatz}, we find 
\begin{align}\label{eq:WaveAnsatzGeneral}
\begin{split}
 \left[\mathcal{H}^{(d)}\right]^{2} = 
 &-v^{2}\sum_{\alpha=1}^{d}\gamma_{\alpha}^{2}\otimes\Delta_{\alpha}
 \\
 &+v^{2}\sum_{\alpha_{1}<\alpha_{2}=2}^{d}\{\gamma_{\alpha_{1}},\gamma_{\alpha_{2}}\}\otimes\mathcal{O}_{\alpha_{1}}\mathcal{O}_{\alpha_{2}},
\end{split}
\end{align}
where we have used the definition of the operators $\mathcal{O}_{\alpha}$ from Eq.~\eqref{eq:SqrtMinusLaplace} as well as the fact that they commute when acting on different dimensions,
\begin{align}
 \left[\mathcal{O}_{\eta},\mathcal{O}_{\nu}\right] = 0.
\end{align}
For Eq.~\eqref{eq:WaveAnsatzGeneral} to satisfy Eq.~\eqref{eq:WaveHamiltonianAnsatzRequired}, the operators $\{\gamma_{\alpha}\}$ must satisfy the anticommutation relations
\begin{align}\label{eq:AnticommutationRelations}
 \{\gamma_{\nu},\gamma_{\eta}\} = 2\delta_{\nu,\eta}\mathds{1},
\end{align}
where $\delta_{\nu,\eta}$ is the Kronecker delta. 
Operators that satisfy the anticommutation relations Eq.~\eqref{eq:AnticommutationRelations} are commonly used in fermion-to-qubit transformations, and many representations of $\{\gamma_{\alpha}\}$ in terms of Pauli strings are known.
For example, it is straightforward to construct an appropriate set of $\{\gamma_{\alpha}\}$ using the ternary-tree-based method~\cite{JiEtAl20}; for additional details, we refer the reader to Sec.~\ref{sec:dDimensionalWaveEquation}.

Having found an appropriate representation of $\mathcal{H}^{(d)}$ we are left with the task of embedding the discretized scalar field $\ket{f}$ and its time derivative $\ket{\partial_{t}f}$ in the quantum state $\ket{\psi}$ associated to the Schr\"{o}dinger equation \eqref{eq:WaveSchrodinger}.
Let $\ket{\psi}$ have the block structure
\begin{align}\label{eq:WaveDirectSumStructure}
 \ket{\psi} = \sum_{\zeta=0}^{K-1}\ket{\zeta}\otimes\ket{\psi_{\zeta}}.
\end{align}
By selecting a basis vector $\ket{\zeta}$ and choosing a particular representation of the operators $\{\gamma_{\alpha}\}$, we find the $\{\ket{\psi_{\zeta}}\}$ that satisfy Eq.~\eqref{eq:WaveSchrodingerStateA}.
For example, by choosing $\ket{\zeta}=\ket{0}$ and 
\begin{align}
 \mathcal{H}^{(d)}=v\sum_{\zeta,\nu}\sum_{\alpha=1}^{d}\left(\gamma_{\alpha}\right)_{\zeta,\nu}\ket{\zeta}\bra{\nu}\otimes\mathcal{O}_{\alpha},
\end{align}
the state $\ket{\psi}$ has the form
\begin{align}\label{eq:WaveStateDefined}
 \sum_{\zeta=0}^{K-1}\ket{\zeta}\otimes\ket{\psi_{\zeta}} \propto \ket{0}\otimes\ket{\partial_{t}f}-\text{i} v \sum_{\zeta=0}^{K-1}\sum_{\alpha=1}^{d}\left(\gamma_{\alpha}\right)_{\zeta,0}\ket{\zeta}\otimes \mathcal{O}_{\alpha}\ket{f}.
\end{align}

An alternative construction is achieved by applying the transformation $\text{i}\left(\mathcal{H}^{(d)}\right)^{-1}$ to both sides of the wave equation Eq.~\eqref{eq:SubspaceWaveEquationA}.
The resulting equation
\begin{align}\label{eq:SubspaceWaveEquationB}
 \ket{\zeta}\otimes\ket{\partial_{t}f}+\text{i}\left(\mathcal{H}^{(d)}\right)^{-1}&\ket{\zeta}\otimes\ket{\partial_{t}^{2}f} =\nonumber\\ &-\text{i}\mathcal{H}^{(d)}\ket{\zeta}
\otimes\ket{f} + \ket{\zeta}\otimes\ket{\partial_{t}f}.
\end{align}
can be recast as the Schr\"{o}dinger equation \eqref{eq:WaveSchrodinger}, where, in this case, the state $\ket{\psi}$ is given by
\begin{align}\label{eq:WaveSchrodingerStateB}
 \ket{\psi} \propto \ket{\zeta}\otimes \ket{f}+\text{i}\left(\mathcal{H}^{(d)}\right)^{-1}\ket{\zeta}\otimes\ket{\partial_{t}f}.
\end{align}
For a particular choice of the basis vector $\ket{\zeta}$ and $\mathcal{H}^{(d)}$ a set of equations specifying $\ket{\psi}$ analogous to Eq.~\eqref{eq:WaveStateDefined} can be found.

\begin{figure}
\centering
\includegraphics{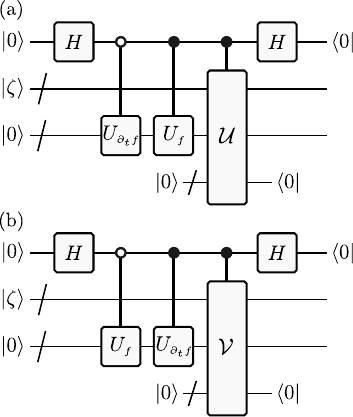}
\caption{\label{fig:WaveEquationStatePrep}
Initial-state preparation circuits for the $d$-dimensional wave equation.
Given efficient circuits to prepare $|f\rangle = U_{f} |\boldsymbol{0}\rangle$ and $|\partial_{t}f\rangle = U_{\partial_{t} f} |\boldsymbol{0}\rangle$, we prepare the state $|\psi\rangle$ in Eqs.~\eqref{eq:WaveSchrodingerStateA} and~\eqref{eq:WaveSchrodingerStateB} using the quantum circuit (a) and (b), respectively.
In (a) the operator $\mathcal{U}$ transforms $|\zeta\rangle \otimes |f\rangle$ into $-\text{i} \mathcal{H}^{(d)} |\zeta\rangle \otimes |f\rangle$.
In (b) the operator $\mathcal{V}$ transforms $|\zeta\rangle \otimes |\partial_{t}f\rangle$ into $\text{i} \left(\mathcal{H}^{(d)}\right)^{-1} |\zeta\rangle \otimes |\partial_{t}f\rangle$.
}
\end{figure}

In the following, we describe one approach to prepare the initial state $|\psi\rangle$ of Eqs.~\eqref{eq:WaveSchrodingerStateA} and~\eqref{eq:WaveSchrodingerStateB}, respectively; see also~\cite{CoJoOs19, BaEtAl23}.
We assume that quantum circuits for the preparation of $|f\rangle$ and $|\partial_{t}f\rangle$ are given.
Figure~\ref{fig:WaveEquationStatePrep} shows our quantum circuit constructions.
A single ancilla qubit is used to sum the two parts of the initial state.
The operators $\mathcal{U}$ and $\mathcal{V}$ implement $-\text{i}\mathcal{H}^{(d)}$ and $\text{i}\left(H^{(d)}\right)^{-1}$, respectively.
The operator $\mathcal{U}$ can be implemented by block-encoding $\mathcal{H}^{(d)}$, which can be efficiently block-encoded as explained in Sec.~\ref{sec:dDimensionalWaveEquation}.
The operator $\mathcal{V}$ can be realized using a quantum linear systems algorithm (QLSA)~\cite{MoEtAl25}.
If $\mathcal{H}^{(d)}$ is not invertible, we use its pseudo-inverse $\left(\mathcal{H}^{(d)}\right)^{+}$ to prepare $\text{i} \left(\mathcal{H}^{(d)}\right)^{+} |\zeta\rangle \otimes |\partial_{t}f\rangle$.
Note that this restricts the class of solutions that can be simulated to those with initial states $|\zeta\rangle \otimes |\partial_{t}f\rangle$ from outside the kernel of $\mathcal{H}^{(d)}$.
The computational cost of preparing $\text{i} \left(\mathcal{H}^{(d)}\right)^{-1} |\zeta\rangle \otimes |\partial_{t}f\rangle$ depends on the QLSA used and, for efficient QLSAs, scales polynomially with the condition number of $\mathcal{H}^{(d)}$.
We obtain an upper bound for the condition number of $\mathcal{H}^{(d)}$ from the one of $\left(\mathcal{H}^{(d)}\right)^{2} = -v^{2} \mathds{1} \otimes \Delta^{(d)}$ which is upper-bounded by $v^{2} d N^{2}$.
Therefore there exists a QLSA that can efficiently prepare $\mathcal{V}$.

\subsection{One spatial dimension}

The goal of this section is to construct quantum circuits for the 1D $(d=1)$ isotropic acoustic wave equation
\begin{align}\label{eq:1DWE}
 \partial_{t}^{2}\ket{f} = v^{2}\Delta^{(1)}\ket{f} = v^{2}\Delta_{1}\ket{f}.
\end{align}
The associated Schr\"{o}dinger equation given by Eq.~\eqref{eq:WaveSchrodinger} is
\begin{align}\label{eq:1DSE}
 \partial_{t}\ket{\psi}= -\text{i}\mathcal{H}^{(1)}\ket{\psi},
\end{align}
and one possible choice of the Hamiltonian satisfying Eq.~\eqref{eq:WaveHamiltonianAnsatz} is
\begin{align}\label{eq:1DHWE}
 \mathcal{H}^{(1)} =vX\otimes\mathcal{O}_{1} =  v\begin{pmatrix}
  0 & \mathcal{O}_{1}
  \\
  \mathcal{O}_{1} & 0
 \end{pmatrix},
\end{align}
where we have chosen $\gamma_{1}=X$, the Pauli $X$ matrix.
We consider two different encodings of $\ket{f}$ and $\ket{\partial_{t}f}$ into the normalized state $\ket{\psi}$.
In particular using Eq.~\eqref{eq:WaveSchrodingerStateA} with $\ket{\zeta}=\ket{0}$ we have the encoding
\begin{align}\label{eq:1DEncoding1}
 \ket{\psi} \propto \ket{0}\otimes\ket{\partial_{t}f}  -\text{i}vX\ket{0}\otimes \mathcal{O}_{1}\ket{f}= \begin{pmatrix}
  \ket{\partial_{t} f}
  \\
  -\im v \mathcal{O}_{1} \ket{f}
 \end{pmatrix}.
\end{align}
Alternatively, using Eq.~\eqref{eq:WaveSchrodingerStateB} with $\ket{\zeta}=\ket{0}$, we have the encoding
\begin{align}\label{eq:1DEncoding2}
 \ket{\psi} \propto \ket{0}\otimes\ket{f}  +\text{i}X\ket{0}\otimes \mathcal{O}^{-1}_{1}\ket{\partial_{t}f}/v= \begin{pmatrix}
  \ket{f}
  \\
  \im \mathcal{O}^{-1}_{1} \ket{\partial_{t}f}/v
 \end{pmatrix}.
\end{align}

For an initial state $\ket{\psi(0)}$, the state $\ket{\psi(\tilde{t})}$ after a time $\tilde{t}$ is given by the solution to Eq.~\eqref{eq:1DSE}
\begin{align}\label{eq:TimeEvolutionOperator1D}
\begin{split}
 \ket{\psi(\tilde{t})} 
 & = \e^{-\im \tilde{t} \mathcal{H}} \ket{\psi(0)}
   = \e^{-\im \tilde{t} vX \otimes \mathcal{O}_{1}} \ket{\psi(0)} 
 \\
 & = (H \otimes \QFT) \e^{-\im t Z \otimes \mathcal{D}_{1}} (H \otimes \QFT^{\dag}) \ket{\psi(0)}.
\end{split}
\end{align}
Here, $H$ is the Hadamard gate, $Z$ is the Pauli $Z$ matrix and we have rescaled time such that $t=\tilde{t}v$.
For periodic boundary conditions we use Eq.~\eqref{eq:EigenvaluesSqrtMinusLaplace} to obtain
\begin{align}\label{eq:TimeEvolutionOperator1DPeriodic}
 \ket{\psi(t)} 
 = (H \otimes \QFT) \e^{-\im 2 t N Z \otimes \sin(\pi \hat{k} / N)} (H \otimes \QFT^{\dag}) \ket{\psi(0)}.
\end{align}
We proceed by rewriting the operator in the argument of the exponential of Eq.~\eqref{eq:TimeEvolutionOperator1DPeriodic} as $Z\otimes \sin(\pi \hat{k}/N)=\sin(\pi \hat{l}/N)$, where we have introduced the operator $\hat{l}=\hat{k}+N(\mathds{1}-Z)/2$.
This leaves us with the task of implementing the time evolution
\begin{align}\label{eq:TimeEvolutionOperator1DPeriodic_l}
 \ket{\psi(t)} 
 = (H \otimes \QFT) \e^{-\im 2 t N \sin(\pi \hat{l} / N)} (H \otimes \QFT^{\dag}) \ket{\psi(0)}.
\end{align}

By noting the similarity between the operator $\e^{-2\im tN\sin(\pi \hat{l} / N)}$ and that of the incompressible advection equation \eqref{eq:TimeEvolutionAdvection}, we observe that the same techniques can be used to implement circuits for this one-dimensional wave equation with periodic boundary conditions.

\begin{figure}
\centering
\includegraphics[width=0.9\linewidth]{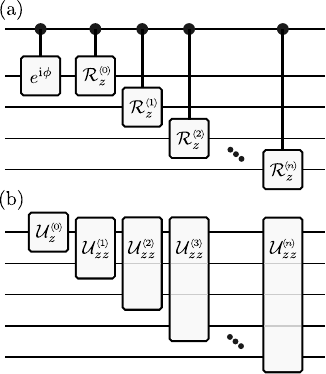}
\caption{\label{fig:wave_equation_1D}
Quantum circuits for the one-dimensional wave equation.
(a) Circuit representation of the operator $U_{\hat{l}}$~\eqref{eq:PlaneWaveBlockEncoding2} where $\phi = \pi(N-1)/(2N)$ and $\mathcal{R}_{z}^{(\zeta)} = \e^{-\im\pi 2^{-\zeta - 1}Z_{\zeta}}$.
(b) Circuit representation of~\eqref{eq:wave_equation_1D_circuit_smooth} where $\mathcal{U}_{z}^{(\zeta)} = \e^{\im \pi t Z_{\zeta}}$ and $\mathcal{U}_{zz}^{(\zeta)} = \e^{\im\pi t 2^{n-\zeta}Z_{0}\otimes Z_{\zeta}}$.
}
\end{figure}

In particular, the approach that makes use of the Jacobi-Anger expansion and is implemented by Fourier-based QSVT scales as $O\left[t N + \log(1/\epsilon)\right]$ for an error $\epsilon$ and is implemented using multiple calls to the operator
\begin{align}\label{eq:PlaneWaveBlockEncoding2}
 U_{\hat{l}} 
 = |0\rangle_{\text{anc}} \langle 0|_{\text{anc}} \otimes \mathds{1} + |1\rangle_{\text{anc}} \langle 1|_{\text{anc}} \otimes \e^{\im\pi \hat{l}/N},
\end{align}
where
\begin{align}\label{PlaneWaveBlockEncoding2}
 \e^{\im\pi\hat{l}/N} = e^{\text{i}\pi/2} e^{-\text{i}\pi/(2N)} \prod_{\beta=0}^{n}\e^{-\im\pi 2^{-\beta-1}Z_{\beta}}
\end{align}
and we have used the definition
\begin{align}
 \hat{l} = \frac{N-1}{2} \mathds{1} - \sum_{\beta = 0}^{n}2^{n-\beta-1}Z_{\beta}.
\end{align}
The quantum circuit representation of Eq.~\eqref{eq:PlaneWaveBlockEncoding2} is shown in Fig.~\ref{fig:wave_equation_1D}~(a).
Furthermore, analogous to the incompressible advection equation, it is possible to construct a circuit scaling as $O(N)$ that implements the time evolution by making use of the discrete Fourier transform to express
\begin{align}
 \e^{-\im2tN\sin(\pi \hat{l}/N)} = \sum_{\zeta=-N/2}^{N/2-1}c_{\zeta}\e^{-\im2\pi\zeta\hat{l}/N},
\end{align}
where
\begin{align}
 c_{\zeta} = \frac{1}{N}\sum_{\tilde{l}=-N/2}^{N/2-1}\e^{-\im 2 t N \sin(\pi \tilde{l} / N)}\e^{-\im2\pi\tilde{l}\zeta/N}.
\end{align}
Taking into account the circuit depth $n$ of Eq.~\eqref{eq:PlaneWaveBlockEncoding2} we find that Eq.~\eqref{eq:TimeEvolutionOperator1DPeriodic} has a circuit representation of depth scaling as $O(n\min(N,tN +\log(1/\epsilon)))$.

Shallower circuits can be derived for smooth initial functions $f$ and $\partial_{t}f$ such that the initial state $\ket{\psi(0)}$ has significant overlaps $|\braket{\tilde{k}}{\psi(0)}|^{2}$ with only small-magnitude wavenumber basis states satisfying $|\tilde{k}|/N \ll 1$.
To derive the associated circuits we consider Eq.~\eqref{eq:TimeEvolutionOperator1DPeriodic} and use the small-angle approximation $\sin(\pi\hat{k}/N)\approx \pi \hat{k}/N$, such that
\begin{align}\label{eq:wave_equation_1D_circuit_smooth}
 \e^{-\im2tNZ\otimes\sin(\pi\hat{k}/N)} &\approx \e^{-\im2\pi t Z\otimes \hat{k}}\nonumber\\
 &= \e^{\im\pi t Z_{0}}\prod_{\beta=1}^{n}\e^{\im\pi t 2^{n-\beta}Z_{0}\otimes Z_{\beta}}.
\end{align}
A circuit representation of the operator \eqref{eq:wave_equation_1D_circuit_smooth} is given in Fig.~\ref{fig:wave_equation_1D}~(b) and is composed of single- and two-qubit gates.

\subsection{Two spatial dimensions}

The objective of this section is to construct quantum circuits for the 2D $(d=2)$ isotropic acoustic wave equation
\begin{align}\label{eq:2DWE}
 \partial_{t}^{2}\ket{f} = v^{2}\Delta^{(2)}\ket{f}=v^{2}\left( \Delta_{1} + \Delta_{2} \right)\ket{f},
\end{align}
where the discretized scalar function $\ket{f}$ represents a two-dimensional wavefield.
The associated Schr\"{o}dinger equation given by Eq.~\eqref{eq:WaveSchrodinger} is
\begin{align}\label{eq:2DSE}
 \partial_{t} \ket{\psi} = -\im \mathcal{H}^{(2)} \ket{\psi}.
\end{align}
One possible choice of Hamiltonian satisfying Eqs.~\eqref{eq:WaveHamiltonianAnsatz} and~\eqref{eq:AnticommutationRelations} is
\begin{align}\label{eq:2DHWE}
 \mathcal{H}^{(2)}/v = X \otimes \mathcal{O}_{1} + Y \otimes \mathcal{O}_{2} =\begin{pmatrix}
  0 & \mathcal{O}_{1} - \im \mathcal{O}_{2}\\
  \mathcal{O}_{1} + \im \mathcal{O}_{2} & 0
 \end{pmatrix},
\end{align}
where we have chosen $\gamma_{1}=X$ and $\gamma_{2}=Y$, the Pauli $Y$ matrix.
We consider two different encodings of $\ket{f}$ and $\ket{\partial_{t}f}$ into the normalized state $\ket{\psi}$.
The first follows from Eq.~\eqref{eq:WaveSchrodingerStateA} and the choice $\ket{\zeta}=\ket{0}$, giving
\begin{align}\label{eq:2DEncoding1}
\begin{split}
 \ket{\psi}
 &\propto\ket{0}\otimes\ket{\partial_{t}f}-\im \mathcal{H}^{(2)}\ket{0}\otimes \ket{f}
 \\
 &\propto \begin{pmatrix}
  \ket{\partial_{t} f}\\
  -\im v (\mathcal{O}_{1} + \im \mathcal{O}_{2}) \ket{f}
 \end{pmatrix}.
\end{split}
\end{align}
The second encoding follows Eq.~\eqref{eq:WaveSchrodingerStateB} and the choice $\ket{\zeta}=\ket{0}$, giving
\begin{align}\label{eq:2DEncoding2}
\begin{split}
 \ket{\psi}
 &\propto\ket{0}\otimes\ket{f}+\im \left(\mathcal{H}^{(2)}\right)^{-1}\ket{0}\otimes\ket{\partial_{t}f}
 \\
 &\propto \begin{pmatrix}
  \ket{f}\\
  \im(\mathcal{O}_{1} - \im \mathcal{O}_{2})^{-1} \ket{\partial_{t}f}/v
 \end{pmatrix}.
\end{split}
\end{align}
The solution to Eq.~\eqref{eq:2DSE} is
\begin{align}\label{eq:TimeEvolutionOperator2D}
\begin{split}
 \ket{\psi(\tilde{t})} 
 & = \e^{-\im \tilde{t} \mathcal{H}^{(2)}} \ket{\psi(0)}
 \\
 & = \e^{-\im \tilde{t} v (X \otimes \mathcal{O}_{1} + Y \otimes \mathcal{O}_{2})} \ket{\psi(0)}
 \\
 & = \left(\mathds{1} \otimes \QFT^{\otimes 2}\right) \e^{-\im t (X \otimes \mathcal{D}_{1} + Y \otimes \mathcal{D}_{2})} \left(\mathds{1} \otimes \QFT^{\dag\otimes 2}\right) \ket{\psi(0)}
\end{split}
\end{align}
where we have rescaled time $t=\tilde{t}v$.
For periodic boundary conditions, we use Eq.~\eqref{eq:EigenvaluesSqrtMinusLaplace} to obtain
\begin{align}\label{eq:TimeEvolutionOperator2DPeriodic}
\e^{-\im t (X \otimes D_{1} + Y \otimes D_{2})} = \e^{-\im 2 t N \left[X \otimes \sin(\pi \hat{k}_{1} / N)  + Y \otimes \sin(\pi \hat{k}_{2} / N)\right]}
\end{align}
where we assume $N = N_{1} = N_{2}$, i.e.\ each dimension has the same number of grid points.

One way to create quantum gates for the time evolution operator in Eq.~\eqref{eq:TimeEvolutionOperator2DPeriodic} utilizes polynomial-based QSVT.
To proceed, we first consider the operator
\begin{align}\label{eq:2DWEHamiltonian}
 \mathscr{H}^{(2)} 
 = X \otimes \sin\left(\frac{\pi \hat{k}_{1}}{N}\right) + Y \otimes \sin\left(\frac{\pi \hat{k}_{2}}{N}\right).
\end{align}
We then consider the shifted and rescaled Hamiltonian
\begin{align}
 \tilde{\mathscr{H}}^{(2)}
 = \frac{\mathscr{H}^{(2)}-\lambda_{\text{min}}\mathds{1}}{\lambda_{\text{max}}-\lambda_{\text{min}}},
\end{align}
where $\lambda_{\text{max}}=-\lambda_{\text{min}}=\sqrt{2}$, such that $\tilde{\mathscr{H}}^{(2)}$ has an eigenspectrum bounded by $[0,1]$.
We then construct a block encoding of $\tilde{\mathscr{H}}^{(2)}$ up to a constant factor $C^{(2)}$, i.e.\
\begin{align}\label{eq:2DW}
 \mathscr{W}^{(2)} = 
 \begin{pmatrix}
  \tilde{\mathscr{H}}^{(2)}/C^{(2)} & \sqrt{1-\left(\tilde{\mathscr{H}}^{(2)}/C^{(2)}\right)^{2}}
  \\
  \sqrt{1-\left(\tilde{\mathscr{H}}^{(2)}/C^{(2)}\right)^{2}} & -\tilde{\mathscr{H}}^{(2)}/C^{(2)}
 \end{pmatrix},
\end{align}
where the resulting block-encoded Hamiltonian has a spectrum bounded by $[0,1/C^{(2)}]$.

\begin{figure}
\centering
\includegraphics[width=0.95\linewidth]{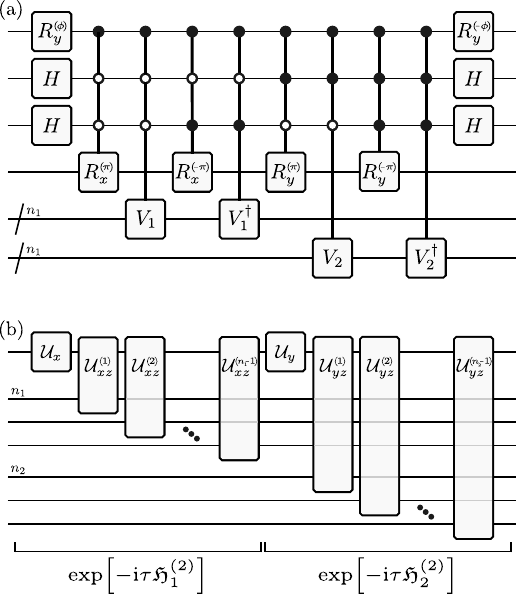}
\caption{\label{fig:wave_equation_2D}
Circuits for the 2D wave equation.
(a) Circuit representation of the operator $\mathscr{W}^{(2)}$~\eqref{eq:2DW}.
Here $H$ is the Hadamard gate, $R_{y}^{(\theta)} = \e^{-\im\theta Y/2}$, $R_{x}^{(\theta)} = \e^{-\im\theta X/2}$, $\phi = 2\cos^{-1}\left(\sqrt{\frac{\sqrt{2}}{2+\sqrt{2}}}\right)$, and $V_{\alpha}=\e^{-\im\pi/(2N_{\alpha})}\prod_{\zeta=0}^{n_{\alpha}-1}\e^{-\im\pi2^{-\zeta-2}Z_{\zeta}}$, where the product runs over the qubit registers $n_1$ and $n_2$ separately.
(b) One layer of a first-order Trotter product formula for a time step $\tau$ approximating the time evolution operator \eqref{eq:Smooth2D}.
Here, the gates $\mathcal{U}_{x} = \e^{\im\pi\tau X}$ and $\mathcal{U}_{y}=\e^{\im\pi\tau Y}$ act on the first qubit while the two-qubit gates $\mathcal{U}^{(\zeta)}_{xz} = \exp(\im\pi\tau2^{n_{1}-\zeta-1}X\otimes Z_{\zeta})$ and $\mathcal{U}^{(\zeta)}_{yz} = \exp(\im\pi\tau2^{n_{2}-\zeta-1}Y\otimes Z_{\zeta})$ act on the first and one qubit from the registers $n_1$ or $n_2$, respectively.
}
\end{figure}

To explicitly construct the block-encoding circuit $\mathscr{W}^{(2)}$, plug the definition of $\hat{k}$ in Eq.~\eqref{eq:HatK} into $\sin(\pi \hat{k}/N)$,
\begin{align}\label{eq:sineCircuit}
\begin{split}
 \sin\left(\frac{\pi \hat{k}}{N}\right) 
 &=\frac{\im}{2}\e^{\im\pi/(2N)}\prod_{\beta=0}^{n-1}\e^{\im\pi2^{-\beta-2}Z_{\beta}}
 \\
 &\quad
 -\frac{\im}{2}\e^{-\im\pi/(2N)}\prod_{\beta=0}^{n-1}\e^{-\im\pi2^{-\beta-2}Z_{\beta}}
\end{split}
\end{align}
Using Eq.~\eqref{eq:sineCircuit} it is straightforward to construct a circuit for the unitary operator $\mathscr{W}^{(2)}$ using the LCU approach as shown in Fig.~\ref{fig:wave_equation_2D}~(a).
By making multiple calls to $\mathscr{W}^{(2)}$, the polynomial-based QSVT approach leads to the nearly-optimal circuit depth $O\left[t N + \log(1/\epsilon)\right]$.

Another way to create a circuit for the time evolution operator in Eq.~\eqref{eq:TimeEvolutionOperator2DPeriodic} assumes that the initial function $f$ and its time derivative $\partial_{t}f$ are smooth.
This allows us to make the small-angle approximation
\begin{align}\label{eq:Smooth2D}
 \e^{-\im 2 t N \left[X \otimes \sin(\pi \hat{k}_{1} / N) + Y \otimes \sin(\pi \hat{k}_{2} / N)\right]} \approx \e^{-\im 2 \pi t (X \otimes \hat{k}_{1} + Y \otimes \hat{k}_{2})}.
\end{align}
The resulting operator corresponds to unitary time evolution with Hamiltonian $\mathfrak{H}^{(2)} = \mathfrak{H}^{(2)}_{1} + \mathfrak{H}^{(2)}_{2}$ where $\mathfrak{H}^{(2)}_{1} = 2 \pi X \otimes \hat{k}_{1}$ and $\mathfrak{H}^{(2)}_{2} = 2 \pi Y \otimes \hat{k}_{2}$.
The corresponding time evolution can be approximated using a Trotter product formula.
In Fig.~\ref{fig:wave_equation_2D}~(b) we show one layer of a first-order Trotter product formula with time step $\tau$, where the unitary operators $\e^{-\im \tau \mathfrak{H}^{(2)}_{1}}$ and $\e^{-\im \tau \mathfrak{H}^{(2)}_{2}}$ alternate.
Higher-order Trotter product formulas can also be employed using the building blocks provided.

\subsection{Three spatial dimensions}

\begin{figure*}
\centering
\includegraphics[width=0.99\linewidth]{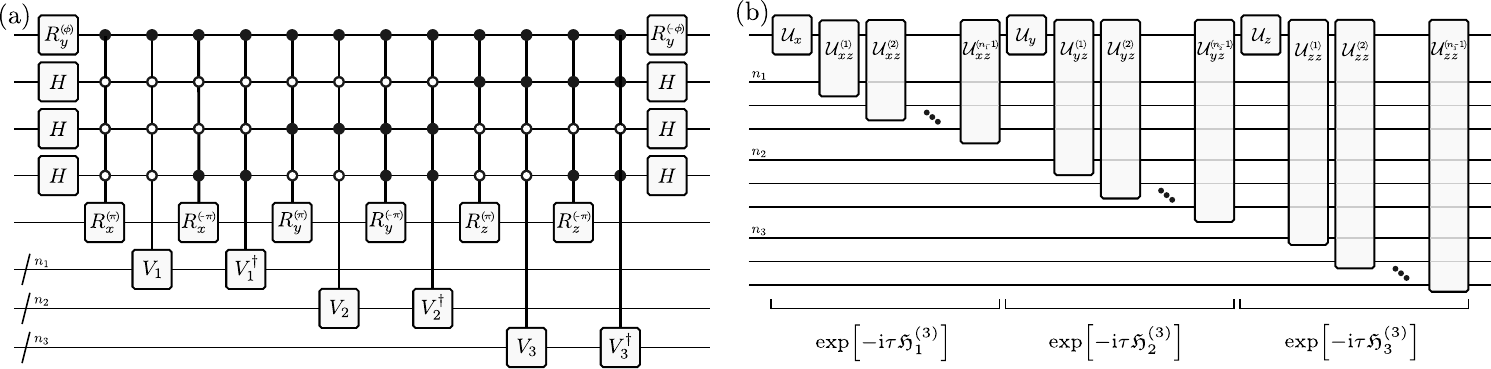}
\caption{\label{fig:wave_equation_3D}
Circuits for the 3D wave equation.
(a) Circuit representation of the block-encoding operator $\mathscr{W}^{(3)}$~\eqref{eq:3DW}.
Here, $H$ is the Hadamard gate,
$R_{x}^{(\theta)}=e^{-\text{i}\theta X/2}$,
$R_{y}^{(\theta)}=e^{-\text{i}\theta Y/2}$, and
$R_{z}^{(\theta)}=e^{-\text{i}\theta Z/2}$.
The rotation angle is $\phi = 2\cos^{-1}\left(\sqrt{\frac{2}{\sqrt{3}}-1}\right)$ and $V_{\alpha}=\e^{-\im\pi/(2N_{\alpha})}\prod_{\zeta=0}^{n_{\alpha}-1}\e^{-\im\pi2^{-\zeta-2}Z_{\zeta}}$, where the product runs over the qubit registers $n_{1}$, $n_{2}$ and $n_{3}$ separately.
(b) One layer of a first-order Trotter product formula approximating the time evolution \eqref{eq:Smooth3D}.
The circuit is composed of one-qubit gates
$\mathcal{U}_{x} = \exp(\text{i}\pi\tau X)$,
$\mathcal{U}_{y} = \exp(\text{i}\pi\tau Y)$,
$\mathcal{U}_{z} = \exp(\text{i}\pi\tau Z)$
that act on the first qubit and two-qubit gates
$\mathcal{U}^{(\zeta)}_{xz} = \exp(\text{i}\pi\tau2^{n_{1}-\zeta-1}X\otimes Z)$,
$\mathcal{U}^{(\zeta)}_{yz} = \exp(\text{i}\pi\tau2^{n_{2}-\zeta-1}Y\otimes Z)$,
and
$\mathcal{U}^{(\zeta)}_{zz} = \exp(\text{i}\pi\tau2^{n_{3}-\zeta-1}Z\otimes Z)$,
which act on the first qubit and one qubit from the registers $n_{1}$, $n_{2}$ and $n_{3}$ respectively.
}
\end{figure*}

The objective of this section is to construct quantum circuits for the 3D $(d=3)$ isotropic acoustic wave equation
\begin{align}\label{eq:3DWE}
 \partial_{t}^{2} \ket{f} = v^{2}\Delta^{(3)}\ket{f}=v^{2}\left( \Delta_{1} + \Delta_{2} + \Delta_{3} \right) \ket{f},
\end{align}
where the scalar function $f$ represents a three-dimensional wavefield.
The associated Schr\"{o}dinger equation given by Eq.~\eqref{eq:WaveSchrodinger} is
\begin{align}\label{eq:3DSE}
 \partial_{t} \ket{\psi} = -\im \mathcal{H}^{(3)} \ket{\psi}.
\end{align}
One possible choice of the Hamiltonian satisfying Eq.~\eqref{eq:WaveHamiltonianAnsatz} is
\begin{align}\label{eq:3DHWE}
 \mathcal{H}^{(3)} &=v\left(X\otimes\mathcal{O}_{1}+Y\otimes\mathcal{O} _{2}+Z\otimes\mathcal{O}_{3}\right)\\&=v\begin{pmatrix}
  \mathcal{O}_{3} & \mathcal{O}_{1}-\im\mathcal{O}_{2}
  \\
  \mathcal{O}_{1}+\im\mathcal{O}_{2} & -\mathcal{O}_{3}
 \end{pmatrix},
\end{align}
where we have chosen $\gamma_{1}=X$, $\gamma_{2}=Y$ and $\gamma_{3}=Z$.
We consider two different encodings of $\ket{f}$ and $\ket{\partial_{t}f}$ into the normalized state $\ket{\psi}$.
In particular using Eq.~\eqref{eq:WaveSchrodingerStateA} with $\ket{\zeta}=\ket{0}$ we have the encoding
\begin{align}\label{eq:3DEncoding1}
 \ket{\psi} &\propto \ket{0}\otimes\ket{\partial_{t}f}  -\text{i}\mathcal{H}^{(3)}\ket{0}\otimes \ket{f}\\
 &= \begin{pmatrix}
  \left(\partial_{t} -\im v\mathcal{O}_{3}\right) \ket{f}
  \\
  -\im v \left(\mathcal{O}_{1} + \im\mathcal{O}_{2}\right)\ket{f}
 \end{pmatrix}.
\end{align}
A second encoding follows Eq.~\eqref{eq:WaveSchrodingerStateB} and the choice $\ket{\zeta}=\ket{0}$, giving
\begin{align}\label{eq:3DEncoding2}
 \ket{\psi}&\propto\ket{0}\otimes\ket{f}+\im \left(\mathcal{H}^{(3)}\right)^{-1}\ket{0}\otimes\ket{\partial_{t}f}\\
 &= \begin{pmatrix}
  \ket{f}-\im\left(\Delta^{(3)}\right)^{-1}\mathcal{O}_{3}\ket{\partial_{t}f}/v\\
  -\im\left(\Delta^{(3)}\right)^{-1}(\mathcal{O}_{1} + \im \mathcal{O}_{2}) \ket{\partial_{t}f}/v
 \end{pmatrix}.
\end{align}

The solution to Eq.~\eqref{eq:3DSE} is
\begin{align}\label{eq:TimeEvolutionOperator3D}
\begin{split}
 &\ket{\psi(\tilde{t})}
 \\
 & = \e^{-\im \tilde{t} \mathcal{H}^{(3)}} \ket{\psi(0)}
 \\
 & = \e^{-\im \tilde{t} v (X \otimes \mathcal{O}_{1} + Y \otimes \mathcal{O}_{2} + Z\otimes\mathcal{O}_{3})} \ket{\psi(0)}
 \\
 & = \left(\mathds{1} \otimes \QFT^{\otimes 3}\right) \e^{-\im t (X \otimes \mathcal{D}_{1} + Y \otimes \mathcal{D}_{2} + Z\otimes \mathcal{O}_{3})}\left(\mathds{1} \otimes \QFT^{\dag\otimes3}\right) \ket{\psi(0)}.
\end{split}
\end{align}
where we have rescaled time $t=\tilde{t}v$.
For periodic boundary conditions, we use Eq.~\eqref{eq:EigenvaluesSqrtMinusLaplace} to express
\begin{align}\label{eq:TimeEvolutionOperator3DPeriodic}
 &e^{-\im t \left( X\otimes \mathcal{D}_{1} + Y\otimes\mathcal{D}_{2}+Z\otimes\mathcal{D}_{3}\right)}\\
 &= e^{-\text{i}2tN\left[X\otimes \sin(\pi\hat{k}_{1}/N) + Y\otimes\sin(\pi \hat{k}_{2}/N) + Z\otimes\sin(\pi\hat{k}_{3}/N)\right]}\nonumber,
\end{align}
where we assume the same number of grid points in each dimension, i.e. $N=N_{1}=N_{2}=N_{3}$.
To use polynomial-based QSVT to create quantum gates implementing the time evolution operator Eq.~\eqref{eq:TimeEvolutionOperator3DPeriodic}, we first consider the operator
\begin{align}
 \mathscr{H}^{(3)} = X\otimes\sin\left(\frac{\pi\hat{k}_{1}}{N}\right)+Y\otimes\sin\left(\frac{\pi\hat{k}_{2}}{N}\right)+\nonumber\\
 Z\otimes\sin\left(\frac{\pi\hat{k}_{3}}{N}\right).
\end{align}
We then consider the shifted and rescaled operator
\begin{align}
 \tilde{\mathscr{H}}^{(3)}=\frac{\mathscr{H}^{(3)}-\lambda_{\text{min}}\mathds{1}}{\lambda_{\text{max}}-\lambda_{\text{min}}}
\end{align}
where $\lambda_{\text{max}} = -\lambda_{\text{min}}=\sqrt{3}$ such that $\tilde{\mathscr{H}}^{(3)}$ has a spectrum bounded by $[0,1]$.
We then construct a block encoding of $\tilde{\mathscr{H}}^{(3)}$ up to a constant factor $C^{(3)}$, i.e.
\begin{align}\label{eq:3DW}
 \mathscr{W}^{(3)} = 
 \begin{pmatrix}
  \tilde{\mathscr{H}}^{(3)}/C^{(3)} & \sqrt{1-\left(\tilde{\mathscr{H}}^{(3)}/C^{(3)}\right)^{2}}
  \\
  \sqrt{1-\left(\tilde{\mathscr{H}}^{(3)}/C^{(3)}\right)^{2}} & -\tilde{\mathscr{H}}^{(3)}/C^{(3)}
 \end{pmatrix},
\end{align}
where the resulting block-encoded Hamiltonian has a spectrum bounded by $[0,1/C^{(3)}]$.
To explicitly construct a block-encoding circuit $\mathscr{W}^{(3)}$ we can write $\sin(\pi\hat{k}/N)$ using Eq.~\eqref{eq:sineCircuit} and the LCU approach as illustrated in Fig.~\ref{fig:wave_equation_3D}~(a). 
By making multiple calles to $\mathscr{W}^{(3)}$, the polynomial-based QSVT approach leads to the nearly-optimal circuit depth $O\left[t N + \log(1/\epsilon)\right]$.

An alternative approach to creating the time evolution operator in Eq.~\eqref{eq:TimeEvolutionOperator3DPeriodic} assumes that the initial function $f$ and its time derivative $\partial_{t}f$ are smooth.
This allows us to make the small-angle approximation
\begin{align}\label{eq:Smooth3D}
\begin{split}
 &e^{-\text{i}2tN\left[X\otimes\sin(\pi\hat{k}_{1}/N)+Y\otimes\sin(\pi\hat{k}_{2}/N)+Z\otimes\sin(\pi\hat{k}_{3}/N)\right]}\\
 &\approx e^{-\text{i}2\pi t\left(X\otimes\hat{k}_{1}+Y\otimes\hat{k}_{2}+Z\otimes\hat{k}_{3}\right)}.
\end{split}
\end{align}
The resulting operator corresponds to unitary time evolution with the Hamiltonian $\mathfrak{H}^{(3)} = \mathfrak{H}^{(3)}_{1}+\mathfrak{H}^{(3)}_{2}+\mathfrak{H}^{(3)}_{3}$ where $\mathfrak{H}^{(3)}_{1}=2\pi X\otimes\hat{k}_{1}$, $\mathfrak{H}^{(3)}_{2}=2\pi Y\otimes \hat{k}_{2}$ and $\mathfrak{H}^{(3)}_{3}=2\pi Z\otimes \hat{k}_{3}$.
The corresponding time evolution can be approximated using a Trotter product formula.
In Fig.~\ref{fig:wave_equation_3D}~(b) we show one layer of a first-order Trotter product formula with time step $\tau$, composed of the unitary operators $\e^{-\im \tau \mathfrak{H}^{(3)}_{1}}$, $\e^{-\im \tau \mathfrak{H}^{(3)}_{2}}$, and $\e^{-\im \tau \mathfrak{H}^{(3)}_{3}}$.
Higher-order Trotter product formulas can also be employed using the building blocks provided.

\subsection{$d$ spatial dimensions}
\label{sec:dDimensionalWaveEquation}

We consider the $d$-dimensional isotropic acoustic wave equation~\eqref{eq:WaveEquation} with $N$ gridpoints in each dimension, $N_{\alpha} = N$ for $\alpha = 1, 2, \ldots, d$, and $d > 1$.
Let us choose a representation of the $\{\gamma_{\alpha}\}$ operators using the ternary tree approach of~\cite{JiEtAl20}.
If the dimension $d$ is odd (even), then $d$ mutually anti-commuting operators $\{\gamma_{1}, \gamma_{2}, \ldots, \gamma_{d}\}$ can be represented by $(d-1)/2$-qubit ($d/2$-qubit) Pauli strings that act non-trivially on no more than $\lceil \log_{3}(d)\rceil$($\lceil \log_{3}(d+1)\rceil$) qubits.
To count the number of gates required to block-encode the Hamiltonian, we consider unitary evolution under Pauli strings that act non-trivially on no more than $r$ qubits, controlled by $q$ qubits.
We denote these gates by
\begin{align}\label{eq:WaveBlockEncodingGateForm}
 \Lambda_{q}[\exp(\im\theta P_{r})]
\end{align}
and count the total number of gates in the block-encoding circuit:
There are $2d$ $(1+\lceil \log_{2}(2d)\rceil)$-controlled rotations $\exp(\text{i}\pi P_{r})$ where the $P_{r}$ are Pauli strings of weight no more than $\lceil \log_{3}(d) \rceil$ ($\lceil\log_{3}(d+1)\rceil$) for $d$ being odd (even).
In addition, there are $2 d n$ $(1+\lceil \log_{2}(2d)\rceil)$-controlled single-qubit Pauli-$Z$ rotation gates.
Finally, there are $2 d$ $(1+\lceil \log_{2}(2d)\rceil)$-controlled phase gates.
The number and form of these gates are tabulated in the first and second column of Tab.~\ref{tab:dWaveBlockEncoding}.

\begin{table}
\caption{\label{tab:dWaveBlockEncoding}
Gates involved in block-encoding of the $d$-dimensional isotropic acoustic wave equation and gate complexity in terms of CNOTs.
}
\begin{ruledtabular}
\begin{tabular}{lll}
 number & gates & CNOTs\\
 \midrule
 odd $d$&&\\
 \cmidrule{1-1}
 $2d$& $\Lambda_{(1+\lceil \log_{2}(2d)\rceil)}[\exp(\im\theta P_{\lceil \log_{3}(d)\rceil})]$ & $O(d\log(d))$\\
 $2dn$& $\Lambda_{(1+\lceil \log_{2}(2d)\rceil)}[\exp(\im\theta P_{1})]$&$O(dn\log(d))$\\
 $2d$& $\Lambda_{(1+\lceil \log_{2}(2d)\rceil)}[\exp(\im\theta)]$&$O(d\log^{2}(d))$\\
 \midrule
 even $d$\\
 \cmidrule{1-1}
 $2d$&$\Lambda_{(1+\lceil \log_{2}(2d)\rceil)}[\exp(\im\theta P_{\lceil \log_{3}(d+1)\rceil})]$ &$O(d\log(d))$\\
 $2dn$& $\Lambda_{(1+\lceil \log_{2}(2d)\rceil)}[\exp(\im\theta P_{1})]$ &$O(dn\log(d))$\\
 $2d$& $\Lambda_{(1+\lceil \log_{2}(2d)\rceil)}[\exp(\im\theta)]$ &$O(d\log^{2}(d))$\\
\end{tabular}
\end{ruledtabular}
\end{table}

It is well-known that weight-$r$ Pauli rotation gates can be decomposed into $O(r)$ CNOTs and single-qubit gates~\cite{NiCh10, ClBaCu21}.
By doing so, the $r$-qubit operation controlled by $q$ qubits is transformed to a single-qubit $R_{z}$ controlled by $q$ qubits, $O(r)$ CNOT gates and additional single-qubit gates.
The single $R_{z}$ gate controlled by $q$ qubits can be implemented with $O(q)$ CNOT gates and additional single-qubit gates~\cite{BaEtAl95, ZiBo25}.
A $q$-qubit-controlled phase gate can be realized using $O\left(q^{2}\right)$ CNOTs~\cite{BaEtAl95, NiCh10}.
We note that, with the help of one ancilla qubit, a $q$-qubit-controlled phase gate can be realized using $O(q)$ CNOTs~\cite{BaEtAl95, RoDuDe25}, but for the sake of simplicity we do not make use of additional ancilla qubits.
The resulting gate complexities in terms of CNOTs are tabulated in the third column of Table~\ref{tab:dWaveBlockEncoding}.

We conclude that the circuit depth for block-encoding the desired Hamiltonian scales as $\tilde{O}(d n)$.
Therefore the total circuit depth of the polynomial-based QSVT approach for realizing the desired time evolution operator is bounded by $\tilde{O}\left(d n \left[t N + \log(1/\epsilon)\right]\right)$.

Let us now determine the circuit depth of the Trotter circuit for smooth initial conditions.
First, we derive the total gate count per Trotter layer.
There are $d$ rotation gates $\exp(\text{i}\pi\tau P_{\alpha})$ and $d n$ rotation gates $\exp(\text{i} \phi P_{\alpha} \otimes Z)$, where the $P_{\alpha}$ are Pauli strings of weight no more than $\lceil \log_{3}(d) \rceil$ ($\lceil \log_{3}(d+1) \rceil$) for $d$ being odd (even).
Each exponential of such a Pauli string can be realized via $O\left[\log(d)\right]$ two-qubit gates~\cite{NiCh10, ClBaCu21}.
Therefore, the total circuit depth per Trotter layer can be bounded by $O\left[d n \log(d)\right]$.
Next, we calculate the error per Trotter layer for time-step size $\tau$.
Using the error upper bound for the first-order Trotter product formula of Eq.~(120) in~\cite{ChEtAl21} and the properties of the $\gamma_{\alpha}$, the Trotter error per layer satisfies
\begin{align}
 \tilde{\epsilon} &\leq \tau^{2} \sum_{\alpha_{1} = 1}^{d} \| \sum_{\alpha_{2} = \alpha_{1}+1}^{d} \gamma_{\alpha_{1}} \gamma_{\alpha_{2}} \otimes \hat{k}_{\alpha_{1}} \otimes \hat{k}_{\alpha_{2}} \|\\
 &\leq \tau^{2} \sum_{\alpha_{1} = 1}^{d} \sum_{\alpha_{2} = \alpha_{1}+1}^{d} \| \gamma_{\alpha_{1}} \gamma_{\alpha_{2}} \otimes \hat{k}_{\alpha_{1}} \otimes \hat{k}_{\alpha_{2}} \|\\
 &\leq \tau^{2} \frac{d (d-1)}{2} k_{\text{max}}^{2}
\end{align}
where $k_{\text{max}}$ is the largest wavenumber required for the smooth function representation.
We obtain the time evolution operator for time $t$ by repeating the Trotter layer $t / \tau$ times which leads an accumulated total error $\epsilon \leq \tilde{\epsilon} t / \tau$.
To achieve a certain desired total error $\epsilon$, it follows that the overall circuit depth scales as $O\left[\log(d) d^{3} k_{\text{max}}^{2} n t^{2} / \epsilon\right]$.

\section{Poisson's equation}
\label{sec:PoissonEquation}

Here we analyze Poisson's equation in $d$ dimensions:
\begin{align}\label{eq:PoissonEquation}
 \left( \frac{\partial^{2}}{\partial x_{1}^{2}} + \frac{\partial^{2}}{\partial x_{2}^{2}} + \ldots + \frac{\partial^{2}}{\partial x_{d}^{2}} \right) f = g
\end{align}
where $f = f(x_{1}, x_{2}, \ldots, x_{d})$ is sought and $g = g(x_{1}, x_{2}, \ldots, x_{d})$ is given.
We utilize~\eqref{eq:SecondDerivative} and obtain the following discretized version of Poisson's equation:
\begin{align}\label{eq:PoissonEquationDiscretized}
 -4 N^{2} \sum_{\alpha=1}^{d} \QFT_{\alpha} \sin^{2}\left(\frac{\pi \hat{k}_{\alpha}}{N}\right) \QFT_{\alpha}^{\dag} |f\rangle = |g\rangle,
\end{align}
which has the formal solution
\begin{align}\label{eq:PoissonSolution}
 |f\rangle = \QFT^{\otimes d} \left[ -4 N^{2} \sum_{\alpha=1}^{d} \sin^{2}\left(\frac{\pi \hat{k}_{\alpha}}{N}\right) \right]^{-1} \left(\QFT^{\dag}\right)^{\otimes d} |g\rangle.
\end{align}
Therefore, to solve Poisson's equation, our goal is to invert the operator
\begin{align}\label{eq:Laplacian}
\begin{split}
 A &= -\sum_{\alpha=1}^{d} \mathcal{D}_{\alpha}^{2}\\
   &= -4 N^{2} \sum_{\alpha=1}^{d} \sin^{2}\left(\frac{\pi \hat{k}_{\alpha}}{N}\right)\\
   &= 2 N^{2} \sum_{\alpha=1}^{d} \left[\cos\left(\frac{2 \pi \hat{k}_{\alpha}}{N}\right) - 1\right].
\end{split}
\end{align}

To create quantum circuits for $A^{-1}$, in the following we present a construction for $d = 1$ and another construction for arbitrary $d$.

\subsection{One dimension}

For Poisson's equation in one dimension, we use the discrete Fourier transform to write
\begin{align}\label{eq:Poisson1DDFT}
 \left[ -4 N^{2} \sin^{2}\left(\frac{\pi \hat{k}}{N}\right) \right]^{-1} = \sum_{\zeta = -N/2}^{N/2 - 1} c_{\zeta} \e^{\im 2 \pi \zeta \hat{k} / N}
\end{align}
where
\begin{align}\label{eq:Poisson1DDFTCoeffs}
 c_{\zeta} = \frac{1}{N} \sum\limits_{\substack{\tilde{k} 
 = -N/2\\\tilde{k} \neq 0}}^{N/2 - 1} \left[ -4 N^{2} \sin^{2}\left(\frac{\pi \tilde{k}}{N}\right) \right]^{-1} \e^{-\im 2 \pi \tilde{k} \zeta / N}.
\end{align}
The sum in~\eqref{eq:Poisson1DDFTCoeffs} excludes the diverging term for $\tilde{k} = 0$ such that Eq.~\eqref{eq:Poisson1DDFT} represents the pseudo-inverse of the one-dimensional Laplace operator.
Fourier-based QSVT with $D = N/2$, $c_{N/2} = 0$ and the operator of Eq.~\eqref{eq:PlaneWaveBlockEncoding} leads to a circuit of depth $O(N)$ representing the right-hand side of Eq.~\eqref{eq:Poisson1DDFT}.

The success probability $p$ for the preparation of the desired operator~\eqref{eq:Poisson1DDFT} scales as $\Omega\left(N^{-4}\right)$.
This follows from the fact that the smallest-magnitude eigenvalue of $\left[ -4 N^{2} \sin^{2}\left(\frac{\pi \hat{k}}{N}\right) \right]^{-1}$ is $-N^{-2}/4$ for $\tilde{k} = -N/2$.
If $|g\rangle$ is a discretized and normalized plane wave of this wavenumber, we obtain the probability $p = N^{-4}/16$ of successfully creating the desired operator via Fourier-based QSVT.
It is important to emphasize that the success probability can be significantly higher for smooth functions $g$.
For example, if $|g\rangle$ has non-vanishing overlaps $|\langle \tilde{k} | g \rangle|^{2}$ only for $|\tilde{k}|/N \ll 1$ up to a maximum wavenumber $|\tilde{k}_{\text{max}}|$, the success probability scales as $\Omega\left(|\tilde{k}_{\text{max}}|^{-4}\right)$.

\subsection{$d$ dimensions}

For Poisson's equation in $d$ dimensions, we note that $\|A\| = 4 d N^{2}$ and the magnitude of the smallest non-vanishing eigenvalue of $A$ is larger than $16$, since $\sin(x) > 2x/\pi$ for $x \in (0, \pi/2)$.
Therefore the condition number of $A$ is $\kappa < d N^{2} / 4 = \tilde{\kappa}$.
To apply QSVT, we introduce the rescaled operator $\tilde{A} = A / \left(4 d N^{2}\right) = A / (16 \tilde{\kappa})$ such that $\|\tilde{A}\| = 1$ is satisfied.
Its smallest non-vanishing eigenvalue in absolute value is larger than $\tilde{\kappa}^{-1}$.

We use Fourier-based QSVT to construct a circuit for the pseudo-inverse of~\eqref{eq:Laplacian}.
According to~\cite{ChKoSo17}, for $\|\tilde{A}\| = 1$, the following Fourier series is $\epsilon$-close to $\tilde{A}^{-1} / \left(16 \tilde{\kappa}\right) = A^{-1}$:
\begin{align}\label{eq:FourierInverse}
 \frac{\im}{16 \tilde{\kappa} \sqrt{2\pi}} \sum_{\zeta = 0}^{G-1} \delta_{y} \sum_{\eta = -K}^{K} \delta_{z}^{2} \eta
 \e^{-(\eta \delta_{z})^{2}/2} \e^{-\im \zeta \eta \delta_{y} \delta_{z} \tilde{A}}
\end{align}
if $G = \Theta\left[\log(1/\epsilon)/\epsilon\right]$, $K = \Theta\left[\tilde{\kappa} \log(1/\epsilon)\right]$, $\delta_{y} = \Theta\left(\tilde{\kappa} \epsilon \left[\log(1/\epsilon)\right]^{-1/2}\right)$, and $\delta_{z} = \Theta\left(\tilde{\kappa}^{-1} \left[\log(1/\epsilon)\right]^{-1/2}\right)$.

The implementation of~\eqref{eq:FourierInverse} via Fourier-based QSVT requires the circuit for the controlled operator $\exp\left[\im \delta_{y} \delta_{z} \mathcal{D}_{\alpha}^{2} / \left(16 \tilde{\kappa}\right)\right]$.
The unitary operator $\exp\left[\im \delta_{y} \delta_{z} \mathcal{D}_{\alpha}^{2} / \left(16 \tilde{\kappa}\right)\right]$ can be approximated by the truncated Jacobi-Anger expansion:
\begin{align}\label{eq:PoissonJA}
\begin{split}
 &\e^{-\im 2 N^{2} \delta_{y} \delta_{z} \left[\cos\left(2 \pi \hat{k}_{\alpha} / N\right) - 1\right]  / \left(16 \tilde{\kappa}\right)}
 \\
 &\approx e^{\im 2 N^{2} \delta_{y} \delta_{z} / \left(16 \tilde{\kappa}\right)}
 \sum_{\nu = -D}^{D} \im^{\nu} J_{\nu}\left[-2 N^{2} \delta_{y} \delta_{z} / \left(16 \tilde{\kappa}\right)\right] \e^{\im \nu 2 \pi \hat{k}_{\alpha} / N}.
\end{split}
\end{align}
We combine the trivial global phases from the $d$ dimensions into one term $\exp\left[\im 2 d N^{2} \delta_{y} \delta_{z} / \left(16 \tilde{\kappa}\right)\right]$.
To achieve overall $O(\epsilon)$ error in~\eqref{eq:FourierInverse}, we need to obtain the truncation error $\epsilon / d G K$ such that the Jacobi-Anger series truncation parameter fulfills
\begin{align}
\label{eq:PoissonDegreeJA}
\begin{split}
 D
 &= \Theta\left(
 \frac{N^{2} \delta_{y} \delta_{z}}{\tilde{\kappa}}
 + \frac{\log(d G K / \epsilon)}{\log\left[\e + \tilde{\kappa} \log(d G K / \epsilon) / \left(N^{2} \delta_{y} \delta_{z}\right)\right]}
 \right)
 \\
 &=
 \Theta\left(\frac{\log(d G K / \epsilon)}{\log\log(d G K / \epsilon) + \log(d / \epsilon) + \log\log(1 / \epsilon)}\right)
 \\
 &= \tilde{O}(1),
\end{split}
\end{align}
where $\tilde{O}(\cdot)$ hides polylogarithmic dependence on parameters.

\begin{figure}
\centering
\includegraphics[width=0.95\linewidth]{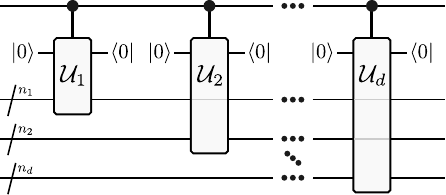}
\caption{\label{fig:Poisson}
Quantum circuit realization of $\exp\left(-\im \delta_{y} \delta_{z} \tilde{A}\right) = \prod_{\alpha = 1}^{d} \exp\left[\im \delta_{y} \delta_{z} \mathcal{D}_{\alpha}^{2} / \left(16 \tilde{\kappa}\right)\right]$ where the $\mathcal{U}_{\alpha}$ are determined by the Fourier-based QSVT construction for Eq.~\eqref{eq:PoissonJA}. 
}
\end{figure}

In $d$ dimensions, the Fourier-based QSVT part of this circuit is applied one dimension after another and all parts are controlled by the same ancilla qubit.
Therefore the total gate complexity of the controlled operator scales as $\tilde{O}(1) \times d = \tilde{O}(d)$.
Note that the ancilla qubit required by Fourier-based QSVT, which is initialized in $|0\rangle$ and must be measured in $|0\rangle$, can be reused by the Fourier-based QSVT circuits in all dimensions as these circuits are run sequentially.
Figure~\ref{fig:Poisson} shows the resulting circuit construction.

We incorporate the controlled operator $\exp\left[\im \delta_{y} \delta_{z} \mathcal{D}_{\alpha}^{2} / \left(16 \tilde{\kappa}\right)\right]$ in another layer of Fourier-based QSVT to realize~\eqref{eq:FourierInverse} which is a Fourier series of $G (2 K + 1)$ terms.
Therefore, the overall gate complexity is proportional to $G (2 K + 1)$ times that of the controlled operator.
Since the latter is $\tilde{O}(d)$, the total gate complexity is
\begin{align}
 G(2K+1) \times \tilde{O}(d)
 = 
 \tilde{O}\left(d \tilde{\kappa} / \epsilon\right)
 \overset{\tilde{\kappa} < d N^{2}}{=}
 \tilde{O}\left(d^{2} N^{2} / \epsilon\right).
\end{align}

The success probability $p$ of the above protocol scales as $\Omega\left(d^{-2} N^{-4}\right)$ as the smallest-magnitude eigenvalue of $A^{-1}$ is $-d^{-1} N^{-2} / 4$.
For sufficiently smooth $|g\rangle$, characterized by the largest-magnitude wavenumber $k_{\text{max}}$ satisfying $|k_{\text{max}}| / N \ll 1$, the success probability scales as $\Omega\left(d^{-2} k_{\text{max}}^{-4}\right)$ and can be much larger than the worst case $p = d^{-2} N^{-4} / 16$.

For smooth $g$ we now derive simpler circuits to solve the Poisson equation.
Using~\eqref{eq:SecondDerivativeSmooth} we approximate $A$ of Eq.~\eqref{eq:Laplacian} by
\begin{align}
 A' = -4 \pi^{2} \sum_{\alpha = 1}^{d} \hat{k}_{\alpha}^{2}.
\end{align}
The smallest-magnitude non-zero eigenvalue of $A'$ is $-4 \pi^{2}$.
We assume that $A'$ acts only on smooth bandwidth-limited functions which are characterized by a maximum wavenumber $k_{\text{max}}$.
In this case, we can restrict the spectrum of $A'$ such that its maximum-magnitude eigenvalue is $-4 \pi^{2} d k_{\text{max}}^{2}$.
The corresponding condition number is $\kappa' = d k_{\text{max}}^{2}$.
We define $\tilde{A}' = A' / \left(4 \pi^{2} \kappa'\right)$ such that $\|\tilde{A}'\| = 1$.
As above, we use the Fourier series approximation of the inverse proposed in~\cite{ChKoSo17} to approximate $\left(\tilde{A}'\right)^{-1}/\left(4 \pi^{2} \kappa'\right) = \left(A'\right)^{-1}$ with error $\epsilon$ via
\begin{align}\label{eq:FourierInverseSmooth}
 \frac{\im}{4 \pi^{2} \kappa' \sqrt{2\pi}} \sum_{\zeta = 0}^{G-1} \delta_{y} \sum_{\eta = -K}^{K} \delta_{z}^{2} \eta
 \e^{-(\eta \delta_{z})^{2}/2} \e^{-\im \zeta \eta \delta_{y} \delta_{z} \tilde{A}'}
\end{align}
where $G = \Theta\left[\log(1/\epsilon)/\epsilon\right]$, $K = \Theta\left[\kappa' \log(1/\epsilon)\right]$, $\delta_{y} = \Theta\left(\kappa' \epsilon \left[\log(1/\epsilon)\right]^{-1/2}\right)$, and $\delta_{z} = \Theta\left(\left(\kappa'\right)^{-1} \left[\log(1/\epsilon)\right]^{-1/2}\right)$.
We construct a circuit for~\eqref{eq:FourierInverseSmooth} using Fourier-based QSVT with the controlled operator
\begin{align}
 e^{-\im \delta_{y} \delta_{z} \tilde{A}'} = \prod_{\alpha = 1}^{d} e^{\im \delta_{y} \delta_{z} \hat{k}_{\alpha}^{2} / \kappa'}
\end{align}
which can be realized in circuit depth of $O\left(d n^{2}\right)$.
Because the Fourier series~\eqref{eq:FourierInverseSmooth} has $\tilde{O}\left(\kappa' / \epsilon\right)$ terms, the overall circuit depth scales as $\tilde{O}\left(d^{2} k_{\text{max}}^{2} n^{2} / \epsilon\right)$.

\section{Discussion}
\label{sec:Discussion}

In this paper we show that, with the help of the QFT, simple quantum circuits can be designed to solve certain high-dimensional PDEs.
The presented circuit constructions can readily be used for other PDEs, such as simple versions of the advection-diffusion or Fokker-Planck equations.
As a next step, it is essential to understand whether the tools developed here can also simplify quantum circuits for the solution of more complicated PDEs that, e.g., are anisotropic, inhomogeneous, subject to complex boundary conditions, or are nonlinear.

In the following we discuss three additional topics for future research.
Firstly, in Sec.~\ref{subsec:InterpCircuits} we present circuit construction protocols that interpolate between our circuits for smooth initial conditions and our QSVT circuits which make no assumptions on smoothness.
Secondly, in Sec.~\ref{subsec:ExpValComp} we show that, if one is only interested in expectation values of observables with respect to the amplitude-encoded PDE solution, in certain cases the required circuits can be further simplified.
Thirdly, as an alternative to the traditional amplitude encoding of PDE solutions on a quantum computer, which is used throughout the paper, we study a different density matrix diagonal encoding in Sec.~\ref{subsec:DensMatDiagEnc}.

\subsection{Interpolating circuit constructions}
\label{subsec:InterpCircuits}

Throughout this paper, we develop quantum circuits for smooth initial conditions and quantum circuits that make no assumptions on the initial conditions.
An interesting question to address is whether there exists a circuit design that interpolates between these two circuit constructions.
One starting point for such an interpolating circuit is the Taylor expansion at $\tilde{k} / N = 0$ of
\begin{align}\label{eq:SineTaylorExpansion}
 N \sin\left(\frac{2 \pi \tilde{k}}{N}\right) 
 = 2 \pi \tilde{k} - \frac{\left(2 \pi \tilde{k}\right)^{3}}{6 N^{2}} + \frac{\left(2 \pi \tilde{k}\right)^{5}}{120 N^{4}} - \ldots
\end{align}
which we truncate at a maximum exponent $\chi$.
The corresponding operator representation reads
\begin{align}\label{eq:SineTaylorOperator}
 N \sin\left(\frac{2 \pi \hat{k}}{N}\right) 
 \approx \sum_{\zeta = 0}^{(\chi-1)/2} \frac{(-1)^{\zeta} (2 \pi)^{2\zeta+1}}{(2\zeta+1)! N^{2\zeta}} \hat{k}^{2\zeta+1}
\end{align}
where we assume that $\chi$ is odd.
As $\chi$ increases in~\eqref{eq:SineTaylorOperator}, the requirement of smooth initial conditions gets systematically more and more relaxed.
Since the right-hand side of Eq.~\eqref{eq:SineTaylorOperator} is a polynomial in $\hat{k}$, its exponential takes on a simple form.
Therefore we can use the approximation of Eq.~\eqref{eq:SineTaylorOperator} to create circuits for the PDEs studied in this paper.

To be specific, let us approximate the central operator of this paper, $\exp\left[-\im t N \sin\left(2 \pi \hat{k} / N\right)\right]$, using~\eqref{eq:SineTaylorOperator}:
\begin{align}\label{eq:SineTaylorExponential}
 \e^{-\im t N \sin\left(2 \pi \hat{k} / N\right)}
 \approx
 \e^{-\im t \sum_{\zeta = 0}^{(\chi-1)/2} \frac{(-1)^{\zeta} (2 \pi)^{2\zeta+1}}{(2\zeta+1)! N^{2\zeta}} \hat{k}^{2\zeta+1}}.
\end{align}
One possibility to prepare a circuit for the right-hand side of Eq.~\eqref{eq:SineTaylorExponential} utilizes the fact that $\hat{k}$ is a sum of commuting operators (see Eq.~\eqref{eq:HatK}).
Due to this fact, the right-hand side of Eq.~\eqref{eq:SineTaylorExponential} is equivalent to
\begin{align}
\prod_{\zeta = 0}^{(\chi-1)/2} e^{-\im t \frac{(-1)^{\zeta} (2 \pi)^{2\zeta+1}}{(2\zeta+1)! N^{2\zeta}} \hat{k}^{2\zeta+1}},
\end{align}
which, for $\chi > 1$, is a product of multi-qubit $Z$ rotation gates that act on at most $\chi$ qubits.
An alternative circuit construction for the right-hand side of Eq.~\eqref{eq:SineTaylorExponential} starts from the Hamiltonian
\begin{align}
 \mathcal{H} = \sum_{\zeta = 0}^{(\chi-1)/2} \frac{(-1)^{\zeta} (2 \pi)^{2\zeta+1}}{(2\zeta+1)! N^{2\zeta}} \hat{k}^{2\zeta+1}
\end{align}
which can be block-encoded using a block-encoding of $\hat{k}$ together with polynomial-based QSVT.
Since $\hat{k}$ is a weighted sum of $O(n)$ unitaries, its block-encoding can be obtained with the help of $O\left[\log(n)\right]$ ancilla qubits and $O(n)$ multi-controlled gates.
It has a circuit depth scaling as $\tilde{O}(n)$ where $\tilde{O}(\cdot)$ ignores polylogarithmic factors.
We utilize the block-encoding of $\hat{k}$ in polynomial-based QSVT to block-encode the polynomial of $\hat{k}$ defining $\mathcal{H}$.
As this is a polynomial of degree $\chi$, the overall circuit depth, including the depth of $\hat{k}$, scales as $\tilde{O}(n \chi)$.
Via the block-encoding of $\mathcal{H}$, we prepare a circuit for $\exp\left(-\im t \mathcal{H}\right)$ using standard QSVT techniques for Hamiltonian simulation.
The resulting circuit is composed of $O\left[t + \log(1/\epsilon)\right]$ block-encodings of $\hat{k}$, for error $\epsilon$, such that the overall circuit depth grows linearly with $\chi$.

\subsection{Expectation value computation}
\label{subsec:ExpValComp}

All circuit constructions of this paper use amplitude encoding to represent the solution of a PDE by a quantum circuit.
If one does not need the solution amplitude-encoded in a circuit but, instead, is only interested in the expectation values of observables, then simpler circuits can be derived.

To see this, we recall that our QSVT-based circuit constructions realize either a Fourier series $\sum_{\zeta = -D}^{D} c_{\zeta} \e^{\im \zeta \mathcal{H}}$ or a polynomial $\sum_{\zeta = 0}^{D} c_{\zeta} \mathcal{H}^{\zeta}$ of a certain Hermitian operator $\mathcal{H}$.
When the PDE solution in Fourier space corresponds to such a Fourier series or polynomial, the associated quantum state is
\begin{align}
 |f^{(1)}\rangle = \QFT^{\otimes d} \left( \sum_{\zeta = -D}^{D} c_{\zeta} \e^{\im \zeta \mathcal{H}} \right) \left(\QFT^{\dag}\right)^{\otimes d} |f(0)\rangle
\end{align}
or
\begin{align}
 |f^{(2)}\rangle = \QFT^{\otimes d} \left( \sum_{\zeta = 0}^{D} c_{\zeta} \mathcal{H}^{\zeta} \right) \left(\QFT^{\dag}\right)^{\otimes d} |f(0)\rangle,
\end{align}
respectively.
The corresponding expectation value of an observable $B$ is
\begin{align}
 \langle f^{(1)}| B |f^{(1)}\rangle = \sum_{\zeta = -D}^{D} \sum_{\eta = -D}^{D} c_{\zeta}^{*} c_{\eta} C_{\zeta, \eta}^{(1)}
\end{align}
where
\begin{align}\label{eq:SimpleExpVal1}
 &C_{\zeta, \eta}^{(1)} =\nonumber\\
 &\langle f(0)| \QFT^{\otimes d} \e^{-\im \zeta \mathcal{H}} \left(\QFT^{\dag}\right)^{\otimes d} B \QFT^{\otimes d} \e^{\im \eta \mathcal{H}} \left(\QFT^{\dag}\right)^{\otimes d} |f(0)\rangle
\end{align}
for the Fourier-based construction and
\begin{align}
 \langle f^{(2)}| B |f^{(2)}\rangle = \sum_{\zeta = 0}^{D} \sum_{\eta = 0}^{D} c_{\zeta}^{*} c_{\eta} C_{\zeta, \eta}^{(2)}
\end{align}
where
\begin{align}\label{eq:SimpleExpVal2}
 C_{\zeta, \eta}^{(2)} = \langle f(0)| \QFT^{\otimes d} \mathcal{H}^{\zeta} \left(\QFT^{\dag}\right)^{\otimes d} B \QFT^{\otimes d} \mathcal{H}^{\eta} \left(\QFT^{\dag}\right)^{\otimes d} |f(0)\rangle
\end{align}
for the polynomial-based construction.
Therefore, the expectation value can be determined via $O\left(D^{2}\right)$ circuits that compute the quantities $C_{\zeta, \eta}^{(1)}$ or $C_{\zeta, \eta}^{(2)}$ of Eqs.~\eqref{eq:SimpleExpVal1} and~\eqref{eq:SimpleExpVal2}.
These circuits can be significantly simpler and easier to realize on quantum hardware than the circuits that are required for the realization of $|f^{(1)}\rangle$ and $|f^{(2)}\rangle$.

\subsection{Density matrix diagonal encoding}
\label{subsec:DensMatDiagEnc}

The postselection success probability of preparing amplitude-encoded solutions to non-unitary evolution, such as for the heat equation, can be very small.
Here we describe an alternative approach that encodes non-negative function values $f_{\ell}$, normalized to fulfill $\sum_{\ell} f_{\ell} = 1$, in the diagonal elements of a density matrix
\begin{align}
 \rho = \sum_{\ell} f_{\ell} |\ell\rangle \langle \ell|.
\end{align}
We show that finite-difference approximations of first and second derivatives can be written as dissipative parts of Lindblad superoperators
\begin{align}\label{eq:LinbladDissipator}
 \mathcal{L}(\rho) = \sum_{\zeta} L_{\zeta} \rho L_{\zeta}^{\dagger} - \frac{1}{2} \{L_{\zeta}^{\dagger} L_{\zeta}, \rho \}
\end{align}
using certain jump operators $L_{\zeta}$.
This enables us to encode the solution to the heat equation on a quantum computer without postselection.

Let us consider a function $f(x)$ in one spatial dimension $x$ and the domain $[0, 1)$ which we discretize into $N = 2^{n}$ equidistant grid points of spacing $s = 1/N$ such that $f_{\ell} = f\left(x^{(\ell)}\right)$ for $x^{(\ell)} = s \ell$ and $\ell = 0, 1, \ldots, N-1$.
We assume periodic boundary conditions such that $f_{-1} = f(-s) = f_{N-1}$ and $f_{N} = f\left(x^{(N-1)}+s\right) = f_{0}$.
We define the jump operators $L_{1} = \sqrt{u} N S$ and $L_{2} = \sqrt{u} N S^{\dagger}$ where $u$ is a non-negative real number, $S |\ell\rangle = |\ell-1\rangle$ for $\ell \in \{1, 2, \ldots, N-1\}$ and $S |0\rangle = |N-1\rangle$.
Then Eq.~\eqref{eq:LinbladDissipator} gives
\begin{align}\label{eq:DissipatorLaplacian}
\begin{split}
 \mathcal{L}(\rho) &= u N^{2} \left( S \rho S^{\dagger} + S^{\dagger} \rho S - 2 \rho \right)\\
 &= u N^{2} \sum_{\ell = 0}^{N-1} \left(f_{\ell+1} - 2 f_{\ell} + f_{\ell-1}\right) |\ell\rangle \langle \ell|\\
 &= u \sum_{\ell = 0}^{N-1} \frac{f\left(x^{(\ell)}+s\right) - 2 f\left(x^{(\ell)}\right) + f\left(x^{(\ell)}-s\right)}{s^{2}} |\ell\rangle \langle \ell|
\end{split}
\end{align}
which leaves the central difference approximation of the one-dimensional Laplacian~\eqref{eq:FiniteDifferenceLaplace} in the diagonal entries of $\mathcal{L}(\rho)$.
The heat equation~\eqref{eq:HeatEquation} in $d = 1$ discretized dimension is then simply given by the purely dissipative Lindblad master equation
\begin{align}
 \frac{\partial \rho}{\partial t} = \mathcal{L}(\rho).
\end{align}

To implement the dissipative evolution $e^{t \mathcal{L}}(\rho)$ on a quantum computer, for each jump operator $L_{\zeta}$ we use one ancilla qubit and the operator
\begin{align}
 K_{\zeta} = |1\rangle \langle 0|_{\text{anc}} \otimes L_{\zeta} + |0\rangle \langle 1|_{\text{anc}} \otimes L_{\zeta}^{\dagger}
\end{align}
that acts on the dilated space.
By applying the unitary operator $\exp(-\text{i} \sqrt{\tau} K_{\zeta})$ for a small time step $\tau$ and then tracing over the ancilla qubit, we approximate the part of $e^{\tau \mathcal{L}}(\rho)$ corresponding to $L_{\zeta}$:
\begin{align}\label{eq:LindbladFirstOrder}
\begin{split}
 &\text{tr}_{\text{anc}} \left[ e^{-\text{i} \sqrt{\tau} K_{\zeta}} \left( |0\rangle \langle 0|_{\text{anc}} \otimes \rho \right) e^{\text{i} \sqrt{\tau} K_{\zeta}^{\dagger}} \right]\\
 &= \rho + \tau \left( L_{\zeta} \rho L_{\zeta}^{\dagger} - \frac{1}{2} \{L_{\zeta}^{\dagger} L_{\zeta}, \rho \} \right) + O\left(\tau^{2} \|L_{\zeta}\|^{4}\right).
\end{split}
\end{align}
In the case of the heat equation with $L_{1} = \sqrt{u} N S$ and $L_{2} = \sqrt{u} N S^{\dagger}$, it is straightforward to construct a circuit for $\exp(-\text{i} \sqrt{\tau} K_{1})$ and $\exp(-\text{i} \sqrt{\tau} K_{2})$, by rewriting the Hermitian operators $K_{1}$ and $K_{2}$ as:
\begin{align}
 K_{1} &= \sqrt{u} N \left( X_{\text{anc}} \otimes \frac{S + S^{\dagger}}{2} - \text{i} Y_{\text{anc}} \otimes \frac{S - S^{\dagger}}{2} \right),\\
 K_{2} &= \sqrt{u} N \left( X_{\text{anc}} \otimes \frac{S + S^{\dagger}}{2} - \text{i} Y_{\text{anc}} \otimes \frac{S^{\dagger} - S}{2} \right),
\end{align}
and noting that $S$ enjoys a particularly simple diagonal representation in Fourier space, $\mathcal{F}^{\dagger} S \mathcal{F} = \exp\left(\text{i} 2 \pi \hat{k} / N\right)$.
This allows us to write:
\begin{align}
 &\left(\mathds{1} \otimes \mathcal{F}^{\dagger}\right) e^{-\text{i} \sqrt{\tau} K_{1}} \left(\mathds{1} \otimes \mathcal{F}\right)\nonumber\\
 &= e^{-\text{i} \sqrt{u \tau} N \left[X_{\text{anc}} \otimes \cos\left(2 \pi \hat{k} / N\right) + Y_{\text{anc}} \otimes \sin\left(2 \pi \hat{k} / N\right)\right]},\\
 &\left(\mathds{1} \otimes \mathcal{F}^{\dagger}\right) e^{-\text{i} \sqrt{\tau} K_{2}} \left(\mathds{1} \otimes \mathcal{F}\right)\nonumber\\
 &= e^{-\text{i} \sqrt{u \tau} N \left[X_{\text{anc}} \otimes \cos\left(2 \pi \hat{k} / N\right) - Y_{\text{anc}} \otimes \sin\left(2 \pi \hat{k} / N\right)\right]},
\end{align}
which can be expressed as quantum circuits composed of one- and two-qubit gates by adapting the circuit constructions presented in the paper.
Repeated application of the dissipative evolution \eqref{eq:LindbladFirstOrder} can be used to evolve the heat equation up to a desired total evolution time $t$.

The approach is straightforward to generalize to larger spatial dimensions by using the $2 d$ jump operators $\sqrt{u} N S_{\alpha}$ and $\sqrt{u} N S_{\alpha}^{\dagger}$ for $\alpha = 1, 2, \ldots, d$, where $S_{\alpha}$ and $S_{\alpha}^{\dagger}$ act on the qubit register corresponding to dimension $\alpha$.
For a total error $\epsilon$, the computational complexity of this approach to solving the heat equation scales as $O\left(d t^{2} u^{2} N^{4} / \epsilon\right)$ times the cost of $e^{-\text{i} \sqrt{\tau} K_{1}}$ and $e^{-\text{i} \sqrt{\tau} K_{2}}$ which can be efficiently prepared using the tools described in the paper.

Finally, we note that choosing the jump operator $L = \sqrt{r N}S$ for a non-negative real number $r$ gives
\begin{align}
\begin{split}
 \mathcal{L}[\rho] &= r N \left(S \rho S^{\dagger} - \rho \right)\\
 &= r N \sum_{\ell = 0}^{N-1} \left(f_{\ell+1} - f_{\ell}\right) |\ell\rangle \langle \ell|\\
 &= r \sum_{\ell = 0}^{N-1} \frac{f\left(x^{(\ell)}+s\right) - f\left(x^{(\ell)}\right)}{s} |\ell\rangle \langle \ell|
\end{split}
\end{align}
and leaves the forward-difference approximation to the first-order spatial derivative on the diagonal of $\mathcal{L}(\rho)$.
Therefore the density matrix diagonal encoding can also be used for solutions of the incompressible advection equation as well as to solve certain variants of advection-diffusion and Fokker-Planck equations on a quantum computer.

\acknowledgments

We are grateful to Marco Ballarin, Guillermo Preisser and Fr\'ed\'eric Sauvage for helpful discussions and comments related to the manuscript.

\bibliography{references}

\end{document}